\documentclass[nojss]{jss}

\usepackage{orcidlink,thumbpdf,lmodern}

\RequirePackage{amsthm,amsmath,amsfonts,amssymb}

\RequirePackage[authoryear]{natbib}
\RequirePackage{hyperref}
\usepackage{adjustbox}
\usepackage{mathtools}
\usepackage{subcaption}
\mathtoolsset{showonlyrefs=true}
\usepackage{tikz}
\usepackage{bbm}
\usepackage{enumitem}

\newcommand{\tauhat}{\widehat{\tau}}

\newcommand{\NN}{\mathbb{N}}
\newcommand{\PP}{\mathbb{P}}

\newcommand{\RR}{\mathbb{R}}
\newcommand{\EE}{\mathbb{E}}

\newcommand{\N}{\text{N}}
\newcommand\tab[1][1cm]{\hspace*{#1}}
\newcommand\ind{\mathbbm{1}}
\newcommand\normm[1]{\left \lVert #1 \right \rVert}

\usepackage{minted}

\usepackage{framed}

\author{Per August J. Moen\textsuperscript{*}~\orcidlink{0009-0003-9990-8341}\\University of Oslo
   \And Sebastian G. Nielsen\textsuperscript{*}~\orcidlink{0009-0000-3694-2436}\\University of Oslo
   \AND Espen B. Urheim\textsuperscript{*}~\orcidlink{0009-0003-0553-1898}\\University of Oslo
   \And Martin Tveten~\orcidlink{0000-0002-4236-633X}\\Norwegian Computing Center
   \And Ingrid K. Glad~\orcidlink{0000-0002-5156-7109}\\University of Oslo}
   \Plainauthor{P. A. J. Moen, S. G. Nielsen, E. B. Urheim, M. Tveten, I. K. Glad}

\title{\pkg{gridcp}: Fast Online Changepoint Detection in \proglang{Python}}
\Plaintitle{gridcp: Fast Online Changepoint Detection in Python}
\Shorttitle{Fast Online Changepoint Detection in \proglang{Python}}

\Abstract{
Online changepoint detection is the problem of detecting distributional changes in a data stream in real-time. A large body of methodology exists for the offline (fixed-size) setting, but applying these methods online quickly becomes infeasible since the per-observation computational cost and memory consumption typically grow at least linearly with the sample size. A recently proposed grid-based methodology \citep{moen_grid} overcomes this by evaluating an offline test statistic over a sparse geometric grid of split points, with grid points spaced increasingly far apart further in the past. For a wide class of test statistics, this approach keeps update time and memory consumption growing logarithmic in the length of the data stream, while admitting finite-sample guarantees on the detection delay. Building on this methodology, we present \pkg{gridcp}, an open-source \proglang{Python} package that turns offline changepoint tests into efficient online detectors through a single, uniform interface. Users can choose from nine \pkg{Numba}-accelerated built-in tests, spanning changes in the mean, variance, covariance, and regression coefficients, as well as nonparametric tests and generalized likelihood-ratio tests for exponential-family models. Users can also supply their own test, which the package handles identically. For any test, \pkg{gridcp} provides Monte Carlo routines that calibrate the detection threshold to a target false alarm probability or average run length, including data-driven variants when no parametric null model is available. Through simulations and three real-data case studies, we show that calibration is accurate, that runtime scales favorably with both stream length and dimension, and that the complete pipeline, from calibration to deployment, runs efficiently on long real-world streams with only a short detection delay.
}

\Plainkeywords{online, sequential, changepoint detection, Numba, Python}

\Address{
  Per August J. Moen~\orcidlink{0009-0003-9990-8341}\\
  Department of Mathematics\\
  University of Oslo\\
  P.O. Box 1053 Blindern\\
  0316 Oslo, Norway\\
  E-mail: \email{pamoen@math.uio.no}\\
  URL: \url{https://www.mn.uio.no/math/english/people/aca/pamoen/}
\newline\newline
  Sebastian G. Nielsen~\orcidlink{0009-0000-3694-2436}\\
  Department of Mathematics\\
  University of Oslo\\
  P.O. Box 1053 Blindern\\
  0316 Oslo, Norway\\
  E-mail: \email{sebasgni@math.uio.no}\\
  URL: \url{https://www.mn.uio.no/math/english/people/aca/sebasgni/}
\newline\newline
  Espen B. Urheim~\orcidlink{0009-0003-0553-1898}\\
  Department of Mathematics\\
  University of Oslo\\
  P.O. Box 1053 Blindern\\
  0316 Oslo, Norway\\
  E-mail: \email{espenbur@math.uio.no}\\
  URL: \url{https://www.mn.uio.no/math/english/people/aca/espenbur/}
\newline\newline
  Martin Tveten~\orcidlink{0000-0002-4236-633X}\\
  Norwegian Computing Center\\
  P.O. Box 114 Blindern\\
  0314 Oslo, Norway\\
  E-mail: \email{tveten@nr.no}\\
  URL: \url{https://nr.no/en/employees/martin-tveten/}
\newline\newline
  Ingrid K. Glad~\orcidlink{0000-0002-5156-7109}\\
  Department of Mathematics\\
  University of Oslo\\
  P.O. Box 1053 Blindern\\
  0316 Oslo, Norway\\
  E-mail: \email{glad@math.uio.no}\\
  URL: \url{https://www.mn.uio.no/math/english/people/aca/glad/}
}

\begin{document}

\renewcommand{\thefootnote}{\fnsymbol{footnote}}
\footnotetext[1]{These authors contributed equally.}
\renewcommand{\thefootnote}{\arabic{footnote}}

\section{Introduction}\label{sec:intro}



Changepoint detection is the problem of identifying distributional changes in a sequence of data points. In the \emph{offline} setting, a fixed-size data set is analyzed retrospectively, and the goal is typically to estimate both the number and locations of any changepoints. A large body of methodologies have been developed for this setting \citep[see, e.g.,][]{killick2012optimal, fryzlewicz2014wild, kovacs2023seeded}, covering a wide range of change scenarios. In many applications, however, the data are not available all at once, but arrive sequentially over time, and a change must be detected in real-time. This is the \emph{online} setting, where the goal is to flag a distributional change as fast as possible after it occurs, while rarely raising a false alarm when no change has taken place. For example, in medical or industrial condition monitoring \citep[see, e.g.,][]{aminikhanghahi2017survey, letzgus2020change}, a missed or delayed change can have potentially fatal or costly consequences. In finance \citep[][]{banerjee2020}, it can mean acting on a regime shift too late. Since a data stream may in principle be infinitely long, an online detector must keep its per-observation cost and memory use reasonable as data accumulate, while still being able to detect changes reliably.

Ideally, the large body of offline changepoint methodology would lend itself naturally to the online setting. In practice, this is rarely the case. In principle, an offline changepoint test could be applied online by re-running it on the entire sequence each time a new observation arrives. This quickly becomes infeasible, however: the per-observation cost typically grows with the length of the stream, eventually exceeding the time between successive observations, while the memory required to store all observations grows linearly with the sample size. To overcome this, \citet{moen_grid} proposed a grid-based method. Rather than evaluating an offline test statistic at every possible changepoint location, the method restricts attention to a small, dynamically updating grid of candidates, raising an alarm whenever the statistic exceeds a detection threshold at any of them. The grid is spaced geometrically, i.e., dense near the present and increasingly sparse into the past, and recycles changepoint candidates. This recycling keeps the candidate count and required number of sufficient statistics, and with it the per-update cost and memory consumption, logarithmic in the length of the stream, for a wide class of test statistics.  Beyond these computational savings, the method admits finite-sample upper bounds on the detection delay, under mild conditions on the test statistic. These depend on the size of the change and are minimax rate optimal in certain settings \citep[for details, see][]{moen_grid}. Together, these properties allow a broad range of offline tests to be deployed online at low computational cost and with established theoretical guarantees behind them.

In this paper we present \pkg{gridcp}, an open-source \proglang{Python} package that implements the grid-based methodology of \citet{moen_grid}. In \pkg{gridcp}, the user can choose from nine built-in statistical tests covering a variety of change types, including CUSUM tests for changes in the mean, likelihood-ratio tests for changes in variance and covariance, tests for changes in regression coefficients, nonparametric tests, and generalized likelihood-ratio tests for changes in exponential family parameters. Alternatively, the user can supply a custom test through a simple interface, after which the package handles it exactly like the built-in ones. In either case,  \pkg{gridcp} provides Monte Carlo methods for calibrating the detection threshold to a target false alarm probability or average run length, including data-driven variants for settings where no parametric model of the null distribution is available. The computationally intensive parts are compiled with \pkg{Numba} for near-native speed, and the detector uses a functional, immutable-state design that integrates with common streaming frameworks.

A variety of software for changepoint detection already exists in both \proglang{Python} and \proglang{R}. Rather than attempt an exhaustive survey, we focus on the packages most relevant to the online setting and compare them along the two dimensions that matter most there: how cost scales with the length of the stream, and whether the false alarm rate can be calibrated. A more comprehensive overview can be found in the CRAN Task Views for Time Series Analysis~\citep{HyndmanKillick2026}. The packages fall into two broad categories: (i) statistical changepoint detection packages such as \pkg{focus-cpt}, \pkg{ocd}, and \pkg{fast-bocpd}, and (ii) deep learning–based tools such as \pkg{AutoCPD}, \pkg{ScanCP}, and the streaming machine learning library \pkg{River}. The first software package for changepoint detection in \proglang{R} was \pkg{strucchange}~\citep{Zeileis2002}, which focuses on detecting changes in regression coefficients, and features some online tools which monitor cumulative sums of various OLS-based residuals. The \pkg{focus-cpt} package~\citep{romano2026focus} implements the Functional Online CuSUM (FOCuS) family of methods \citep{Romano2023FOCuS,npfocus,ward2025poissonfocus, ward2024constant}, providing likelihood-ratio–based online detection for a range of classical models (Gaussian mean with potential autocorrelated noise, univariate exponential families and nonparametric detection), but without calibration or simulation tools. The \pkg{ocd} package \citep{chen_high-dimensional_2022} focuses on high-dimensional Gaussian mean shifts with identity covariance, implementing the proposed ocd procedure together with several related methods \citep{Mei2010EfficientScalable,XieSiegmund2013SequentialMultiSensor,Chan2017OptimalSequential}; it is restricted to this specific changepoint problem and only provides average run length calibration. The \pkg{fast-bocpd} package offers an efficient implementation of Bayesian Online Changepoint Detection \citep{adams2007bocpd} for a variety of conjugate-exponential family models (Gaussian, Poisson, Bernoulli/Binomial, t-distribution, Gamma) but requires the careful specification of how often one should expect to see a changepoint, lacks calibration methodology, and supports only one pre-specified conjugate prior per changepoint problem.

On the deep learning side, \pkg{AutoCPD} \citep{JieAutoCPD2023} provides a \proglang{Python} framework for training feedforward and convolutional neural networks to perform fixed-window, offline change-or-no-change classification, requiring labeled time series for supervised training and offering no tools for applying the networks to an online setting. The \pkg{ScanCP} package \citep{scanCP} uses feedforward neural networks with sliding windows of lengths $W$ and $2W$ to construct a residual-based test statistic for univariate time series. An online deployment would require continual retraining, which is not implemented. Finally, \pkg{River} \citep{montiel2021river} is a streaming machine learning library that features two online ``drift detection'' methods: ADWIN2 \citep{BifetGavalda2007ADWIN} for detecting mean shifts in bounded data via adaptive windows, and the Page--Hinkley test \citep{page1954continuous} for Gaussian mean changes. These methods are primarily intended for indicating when to retrain models rather than for performing changepoint analysis.

Across these packages, there are several common limitations. \pkg{ocd} and \pkg{river} are only designed for very specific changepoint scenarios; \pkg{AutoCPD} and \pkg{ScanCP} are not designed for fully online operation; none of the packages provide false-alarm calibration, and only the \pkg{ocd} package provides average run length calibration; many methods have a high computational complexity, especially in multivariate settings; multivariate data are often unsupported or just partially supported; and deep learning–based approaches such as \pkg{AutoCPD} additionally require labeled data for training.

Other (mostly offline) changepoint detection packages include \pkg{changepoint}~\citep{Killick2014}, \pkg{fastcpd} \citep{Li2026fastcpd}, \pkg{ruptures}~\citep{truong2020selective}, \pkg{changepoint-cython} \citep{Bruned2023ChangepointCython}, \pkg{pelt} \citep{Versteeg2026Pelt}, \pkg{changepoints} \citep{changepoints2022}, and \pkg{HDCD} \citep{Moen2024HDCD}.

The paper is structured as follows. Section~\ref{sec:background} formally introduces the grid-based methodology that is implemented in  \pkg{gridcp}. Section~\ref{sec:package} summarizes the main design principles behind the package, and provides a simple code example of univariate changepoint detection. Section~\ref{sec:builtins} gives an overview of all 9 built-in score models (changepoint tests) and their associated changepoint problems. Section~\ref{sec:calibration} describes \pkg{gridcp}'s functionality for calibrating any detector to a target false alarm probability or average run length, and we demonstrate its precision empirically. Throughout, the software is illustrated with worked examples on both simulated and real data, with all code provided in an accompanying replication script. 
Additionally, Section~\ref{sec:soundsending} provides an application to a novel real-world data set, where \pkg{gridcp} flags changes in long sensor data streams.  Finally, Section~\ref{sec:discussion} discusses the scope of the methodology, our main design choices, and possible directions for future work.

\paragraph{Notation.} We use the following notation throughout the paper. For a vector $v$, we let $v^\top$ denote its transpose, and we write $v(j)$ for its $j$-th entry. Observations are indexed by time, $Y_1, Y_2, \dots$, with $Y_i \in \mathbb{R}^p$; the subscript denotes the time index and $Y_i(j)$ the $j$-th coordinate of $Y_i$. For integers $a \le b$, we write $a\!\!:\!\!b$ for the index range $\{a, a+1, \dots, b\}$, and let $Y_{a:b} = (Y_a, \dots, Y_b)$ denote the corresponding block of observations. We write $\|\cdot\|_2$ for the Euclidean norm, $\ind\{\cdot\}$ for the indicator function, $I_p$ for the $p \times p$ identity matrix, and $\chi^2(k)$ for the chi-squared distribution with $k$ degrees of freedom. Finally, $\mathcal{O}(\cdot)$ denotes standard big-O asymptotic notation.

\section{Background and methodology}\label{sec:background}

In this section, we summarize the grid-based methodology implemented in \pkg{gridcp}. The companion paper \citep{moen_grid} formally defines the methodology and provides theoretical guarantees. Here we focus on the main ideas needed to understand the software. 

The approach of the methodology to online detection is to scan for changes over a carefully constructed set of split points (i.e., candidate changepoints). This scan is made computationally feasible through two ingredients: a class of test statistics that admit bounded-size running summaries together with constant-time per-split evaluations, and a sparse geometric-like grid of split points. Together, these ingredients allow a broad family of offline changepoint tests to be repurposed for online monitoring with storage and update costs that scale logarithmically with the sample size. 

\subsection{Problem setup}

We observe a sequence of $p$-dimensional data points $Y_1, Y_2, \ldots$ arriving sequentially. At some unknown time index $\tau \in \{2,3,\ldots\}\cup\{\infty\}$, the data-generating distribution changes: $Y_i \sim P_1$ for $i < \tau$ and $Y_i \sim P_2$ for $i \geq \tau$. Thus, $\tau$ is the index of the first post-change observation.\footnote{In the statistical changepoint literature, the changepoint is typically defined as the \emph{last} pre-change observation index. We have opted for the former to align with established \proglang{Python} conventions.} Given $\tau$, we write $\PP_{\tau}$ and $\EE_{\tau}$ for probability and expectation under this model, with $\tau = \infty$ corresponding to the no-change regime. An online detector is a stopping rule that raises an alarm once sufficient evidence for a change has accumulated, and we denote by $\tauhat$ the first time index at which an alarm is raised. The overall goal is to detect the change quickly when $\tau < \infty$, while controlling the frequency of false alarms when $\tau = \infty$. We measure false alarm frequency either by the \emph{false alarm probability} $\PP_{\infty}(\tauhat < \infty)$ or by the \emph{average run length} $\EE_{\infty}(\tauhat)$ (ARL). When a changepoint is present $(\tau<\infty$), the main performance measure is the \emph{detection delay}, $\tauhat-\tau$, which is the number of samples from the changepoint to a subsequent alarm. Ideally, the detection delay should be as close to $0$ as possible. Finally, we remark that this single-changepoint setup is a natural building block to deal with multiple changes in practice, since a stopping rule can simply be restarted after an alarm has been raised.

\subsection{Single-split changepoint tests}\label{subsec:test-form}

The methodology of \citet{moen_grid} targets changepoint tests that, for any specified split point $b$, can be written in terms of running summaries of transformed observations. Assume we have observed data $Y_1, \ldots, Y_t$, and let $b\in \{2,3,\ldots, t\}$ be some split point, dividing the sequence into two segments $\{Y_1, \ldots, Y_{b-1}\}$ and $\{Y_b, \ldots, Y_t\}$.
In the software terminology used by \pkg{gridcp}, the numerical value of such a test statistic at this split point is called a \emph{score}. We assume that the score $S_b^{(t)}$ takes the form
\begin{equation}
  S_b^{(t)}
  \;=\;
    f_b^{(t)}\!\bigg(
      \sum_{i=1}^{b-1} h(Y_i),\;
      \sum_{i=b}^{t} h(Y_i)
    \bigg)
  \label{eq:framework-test}
\end{equation}
and that the point-wise test rejects the null hypothesis of no change when $S_b^{(t)} > \lambda \cdot \mathrm{pen}(t)$. Here, $h \colon \RR^p \to \RR^v$ extracts the summary statistics needed by the test, for example $h(y) = y$ for changes in mean and a vectorized outer product for changes in covariance. The map $f_b^{(t)}$ then converts the pre- and post-split summaries into a changepoint score, while $\mathrm{pen}(t)$ is a penalty term and $\lambda > 0$ is the leading rejection threshold constant. If the goal is false alarm probability control over an unbounded horizon, the penalty typically has to grow with time. Meanwhile, for average run length calibration, a constant penalty is often sufficient and preferable.

As an example, the classical CUSUM statistic \citep{wang2020univariate} may be written as
\begin{align}
    \mathrm{CUSUM}_b^{(t)} = \sqrt{\frac{t-b+1}{t(b-1)\vphantom{(t-b)}}}\sum_{i=1}^{b-1} Y_i - \sqrt{\frac{b-1}{t(t-b+1)}}\sum_{i=b}^t Y_i .\label{CUSUMex}
\end{align}
Thus, the squared value of the CUSUM in \eqref{CUSUMex} may be written on the form in \eqref{eq:framework-test}, letting $h$ be the identity map. Moreover, with a suitable choice of $\lambda$, the time-varying penalty $\mathrm{pen}(t) = \log t + \sqrt{\log t}$ can be shown to control the false alarm probability \citep[see][]{moen_grid}, and the constant penalty $\mathrm{pen}(t) = 1$ can be shown to control the average run length \citep[see][]{anote}. 

Returning to the generic score in \eqref{eq:framework-test}, if we write \begin{align}H_j := \sum_{i=1}^j h(Y_i)\label{summary}\end{align} for the sum of the transformed data up to time $j$, then the score in \eqref{eq:framework-test} can be written as $S_b^{(t)} = f_b^{(t)}(H_{b-1}, H_t - H_{b-1})$. This equivalent form is convenient for online computation, since $S_b^{(t)}$ can be computed from time-indexed summaries rather than being re-computed from the raw data. Specifically, $S_b^{(t)}$ can be computed directly from the running sum $H_t$ at the current time $t$, and the running sum up $H_{b-1}$ to the split point $b$. However, for this convenient representation of $S_b^{(t)}$ to admit computationally efficient updating, we impose the assumption that the evaluation of $f_b^{(t)}$ from these summaries, and the evaluation of $\mathrm{pen}(t)$, must be possible in constant time, uniformly over $b$ and $t$. 

As shown in \cite{moen_grid}, several test statistics from the offline changepoint literature may be written on the form of \eqref{eq:framework-test}, such as the CUSUM test shown above. Other tests include tests for changes in regression coefficients, likelihood ratio tests for exponential families, and tests comparing estimated parameters directly \citep[see Section 2.2.1 in][]{moen_grid}. These scores may be computed in $\mathcal{O}(1)$ time for appropriate summaries, and the evaluation of appropriate penalty functions may be computed in $\mathcal{O}(1)$ time as well. Thus, a wide range of established changepoint tests are suitable for the methodology.

To implement a test statistic as a score within \pkg{gridcp}, the user does not provide $h$, the family of maps $\{f_b^{(t)}\}_{b,t}$, and $\mathrm{pen}(t)$ directly to \pkg{gridcp}. Instead, these ingredients are encapsulated in a \emph{score model}. Section~\ref{sec:package} explains how the package represents and updates the corresponding summaries in software. We remark that \pkg{gridcp} ships with multiple built-in scores, including the aforementioned examples above. The built-in scores are described in detail in Section~\ref{sec:builtins}.

\subsection{From single-split tests to grid-based scans}
Because the score in \eqref{eq:framework-test} only targets one single split point $b$, it must be evaluated over multiple split points to retain power against changes that may occur anywhere in the sample. If we choose some set $B^{(t)} \subseteq \{2,3, \ldots,t\}$ of such split points, it is natural to declare an alarm if the score exceeds the threshold at \emph{some} split point $b\in B^{(t)}$, i.e., whenever 
$S_b^{(t)}>\lambda \cdot \mathrm{pen}(t)$ for at least one $b\in B^{(t)}$. Our stopping rule $\tauhat$ therefore equals the smallest sample size $t$ for which 
\begin{align}
  \underset{b \in B^{(t)}}{\max} \  {S}_b^{(t)} > \lambda \cdot \mathrm{pen}(t), \label{overalltest}
\end{align}
or, equivalently, whenever the \textit{penalized score} ${S}_b^{(t)} / \mathrm{pen}(t)$ exceeds $\lambda$ for some $b \in B^{(t)}$. If no such $t$ exists, meaning no change is ever declared, we set $\tauhat=\infty$.

If we take $B^{(t)} = \{2,3, \ldots, t\}$, we recover a full scan over all possible split points, which is statistically natural but quickly becomes very expensive for long data streams. Indeed, a full scan will result in a linear per-update computational cost, even when each score itself can be updated from a fixed-size summary state. A sparse multiscale split point set can reduce that count to logarithmic order, but would not necessarily solve the online memory problem: as $t$ increases, the split points may drift, thus requiring storing all summaries $H_1, H_2, \ldots, H_t$, resulting in a linearly growing memory consumption.

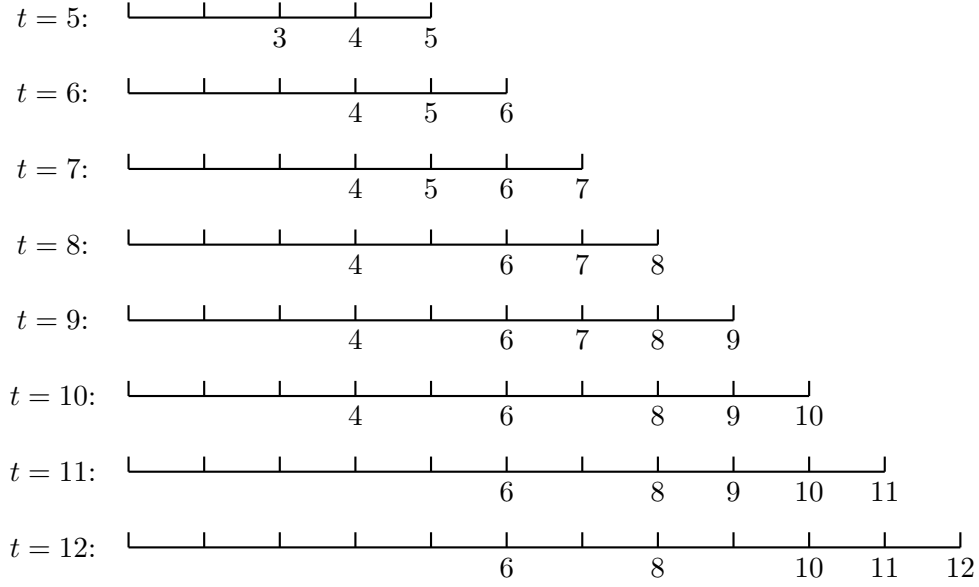
\begin{figure}
\centering
\begin{tikzpicture}

\draw[solid,thick] (1,0.0) -- (5,0.0);
\node at (0,0.0) {$t=5$:};
\draw[thick] (1,0.2) -- ++ (0,-0.2) node[below] {};
\draw[thick] (2,0.2) -- ++ (0,-0.2) node[below] {};
\draw[thick] (3,0.2) -- ++ (0,-0.2) node[below] {$3$};
\draw[thick] (4,0.2) -- ++ (0,-0.2) node[below] {$4$};
\draw[thick] (5,0.2) -- ++ (0,-0.2) node[below] {$5$};

\draw[solid,thick] (1,-1.0) -- (6,-1.0);
\node at (0,-1.0) {$t=6$:};
\draw[thick] (1,-0.8) -- ++ (0,-0.2) node[below] {};
\draw[thick] (2,-0.8) -- ++ (0,-0.2) node[below] {};
\draw[thick] (3,-0.8) -- ++ (0,-0.2) node[below] {};
\draw[thick] (4,-0.8) -- ++ (0,-0.2) node[below] {$4$};
\draw[thick] (5,-0.8) -- ++ (0,-0.2) node[below] {$5$};
\draw[thick] (6,-0.8) -- ++ (0,-0.2) node[below] {$6$};

\draw[solid,thick] (1,-2.0) -- (7,-2.0);
\node at (0,-2.0) {$t=7$:};
\draw[thick] (1,-1.8) -- ++ (0,-0.2) node[below] {};
\draw[thick] (2,-1.8) -- ++ (0,-0.2) node[below] {};
\draw[thick] (3,-1.8) -- ++ (0,-0.2) node[below] {};
\draw[thick] (4,-1.8) -- ++ (0,-0.2) node[below] {$4$};
\draw[thick] (5,-1.8) -- ++ (0,-0.2) node[below] {$5$};
\draw[thick] (6,-1.8) -- ++ (0,-0.2) node[below] {$6$};
\draw[thick] (7,-1.8) -- ++ (0,-0.2) node[below] {$7$};

\draw[solid,thick] (1,-3.0) -- (8,-3.0);
\node at (0,-3.0) {$t=8$:};
\draw[thick] (1,-2.8) -- ++ (0,-0.2) node[below] {};
\draw[thick] (2,-2.8) -- ++ (0,-0.2) node[below] {};
\draw[thick] (3,-2.8) -- ++ (0,-0.2) node[below] {};
\draw[thick] (4,-2.8) -- ++ (0,-0.2) node[below] {$4$};
\draw[thick] (5,-2.8) -- ++ (0,-0.2) node[below] {};
\draw[thick] (6,-2.8) -- ++ (0,-0.2) node[below] {$6$};
\draw[thick] (7,-2.8) -- ++ (0,-0.2) node[below] {$7$};
\draw[thick] (8,-2.8) -- ++ (0,-0.2) node[below] {$8$};

\draw[solid,thick] (1,-4.0) -- (9,-4.0);
\node at (0,-4.0) {$t=9$:};
\draw[thick] (1,-3.8) -- ++ (0,-0.2) node[below] {};
\draw[thick] (2,-3.8) -- ++ (0,-0.2) node[below] {};
\draw[thick] (3,-3.8) -- ++ (0,-0.2) node[below] {};
\draw[thick] (4,-3.8) -- ++ (0,-0.2) node[below] {$4$};
\draw[thick] (5,-3.8) -- ++ (0,-0.2) node[below] {};
\draw[thick] (6,-3.8) -- ++ (0,-0.2) node[below] {$6$};
\draw[thick] (7,-3.8) -- ++ (0,-0.2) node[below] {$7$};
\draw[thick] (8,-3.8) -- ++ (0,-0.2) node[below] {$8$};
\draw[thick] (9,-3.8) -- ++ (0,-0.2) node[below] {$9$};

\draw[solid,thick] (1,-5.0) -- (10,-5.0);
\node at (0,-5.0) {$t=10$:};
\draw[thick] (1,-4.8) -- ++ (0,-0.2) node[below] {};
\draw[thick] (2,-4.8) -- ++ (0,-0.2) node[below] {};
\draw[thick] (3,-4.8) -- ++ (0,-0.2) node[below] {};
\draw[thick] (4,-4.8) -- ++ (0,-0.2) node[below] {$4$};
\draw[thick] (5,-4.8) -- ++ (0,-0.2) node[below] {};
\draw[thick] (6,-4.8) -- ++ (0,-0.2) node[below] {$6$};
\draw[thick] (7,-4.8) -- ++ (0,-0.2) node[below] {};
\draw[thick] (8,-4.8) -- ++ (0,-0.2) node[below] {$8$};
\draw[thick] (9,-4.8) -- ++ (0,-0.2) node[below] {$9$};
\draw[thick] (10,-4.8) -- ++ (0,-0.2) node[below] {$10$};

\draw[solid,thick] (1,-6.0) -- (11,-6.0);
\node at (0,-6.0) {$t=11$:};
\draw[thick] (1,-5.8) -- ++ (0,-0.2) node[below] {};
\draw[thick] (2,-5.8) -- ++ (0,-0.2) node[below] {};
\draw[thick] (3,-5.8) -- ++ (0,-0.2) node[below] {};
\draw[thick] (4,-5.8) -- ++ (0,-0.2) node[below] {};
\draw[thick] (5,-5.8) -- ++ (0,-0.2) node[below] {};
\draw[thick] (6,-5.8) -- ++ (0,-0.2) node[below] {$6$};
\draw[thick] (7,-5.8) -- ++ (0,-0.2) node[below] {};
\draw[thick] (8,-5.8) -- ++ (0,-0.2) node[below] {$8$};
\draw[thick] (9,-5.8) -- ++ (0,-0.2) node[below] {$9$};
\draw[thick] (10,-5.8) -- ++ (0,-0.2) node[below] {$10$};
\draw[thick] (11,-5.8) -- ++ (0,-0.2) node[below] {$11$};

\draw[solid,thick] (1,-7.0) -- (12,-7.0);
\node at (0,-7.0) {$t=12$:};
\draw[thick] (1,-6.8) -- ++ (0,-0.2) node[below] {};
\draw[thick] (2,-6.8) -- ++ (0,-0.2) node[below] {};
\draw[thick] (3,-6.8) -- ++ (0,-0.2) node[below] {};
\draw[thick] (4,-6.8) -- ++ (0,-0.2) node[below] {};
\draw[thick] (5,-6.8) -- ++ (0,-0.2) node[below] {};
\draw[thick] (6,-6.8) -- ++ (0,-0.2) node[below] {$6$};
\draw[thick] (7,-6.8) -- ++ (0,-0.2) node[below] {};
\draw[thick] (8,-6.8) -- ++ (0,-0.2) node[below] {$8$};
\draw[thick] (9,-6.8) -- ++ (0,-0.2) node[below] {};
\draw[thick] (10,-6.8) -- ++ (0,-0.2) node[below] {$10$};
\draw[thick] (11,-6.8) -- ++ (0,-0.2) node[below] {$11$};
\draw[thick] (12,-6.8) -- ++ (0,-0.2) node[below] {$12$};

\end{tikzpicture}
\caption{Split points $B^{(t)}$ for $t = 5, \ldots, 12$. }\label{fig:changelocations}
\end{figure}

The key innovation of \citet{moen_grid} is to propose a particular choice of $B^{(t)}$, called the \emph{dynamic geometric grid}. In the companion paper this construction is expressed through a lag set $G^{(t)}$, parameterizing the split point $b$ via the number of post-change samples $g = t-b+1$. Here, however, it is more natural to work directly with the corresponding set of split points
$B^{(t)} = t +1- G^{(t)} = \{t+1 -g : g \in G^{(t)}\} \subseteq \{2,3, \ldots, t\}$. A plot of the elements of the dynamic geometric grid $B^{(t)}$ is given in Figure~\ref{fig:changelocations}, where split points are indicated by numbered ticks, for $t=5,6,\ldots, 12$.

Expressed in terms of $B^{(t)}$, the essential features of the dynamic geometric grid $B^{(t)}$ are \citep[][Lemma 1]{moen_grid}:
\begin{enumerate}
  \item \textbf{Geometric spacing:} For any (true) lag $d = t+1-\tau \leq t/2$, there exists a split point $b \in B^{(t)}$ whose lag $t+1-b$ satisfies $d/2 \leq t+1-b \leq d$. Hence, $B^{(t)}$ always contains a split point whose lag is within a constant factor of the true lag, which is enough for many changepoint scores to retain most of their signal \citep[see Proposition 3 in][]{moen_grid}.
  \item \textbf{Logarithmic size:} $|B^{(t)}| = O(\log t)$, so only $\mathcal{O}(\log t)$ scores are evaluated at each time step.
  \item \textbf{Recycling:} $B^{(t+1)} \subseteq B^{(t)} \cup \{t+1\}$. Equivalently, when time advances from $t$ to $t+1$, the required summaries $\{H_{b-1}\}_{b \in B^{(t)}}$ are either already stored from the previous step or is the newly created current summary $H_t$.
\end{enumerate}

The three above properties ensure that a detector defined by the stopping rule in \eqref{overalltest}, with $B^{(t)}$ chosen as the dynamic geometric grid, is both statistically effective and computationally scalable. Geometric spacing keeps the grid sparse without sacrificing multiscale coverage, while recycling is what turns logarithmic scan size into logarithmic memory usage as well. \cite{moen_grid} proves these statements formally and shows that, under suitable robustness conditions on the score, scanning over the dynamic grid yields detection delays that are comparable to those of a full scan up to constant factors \citep[][Proposition 3]{moen_grid}. Turning to \pkg{gridcp}, the three properties have the following practical meaning: once a changepoint score can be updated from fixed-size summaries, the package can run it online on the dynamic grid with low computational cost, without storing the entire data history.

\section[The gridcp Package]{The gridcp package}\label{sec:package}


The \pkg{gridcp} package implements the grid-based online changepoint detection methodology described in Section~\ref{sec:background}. It is built on top of NumPy \citep{harris2020array} and designed to be easy to use, computationally efficient, and integrate well with common \proglang{Python}-based streaming frameworks. To install \pkg{gridcp}, run \code{pip install gridcp} in a terminal.

In this section, we first explain the design and general interface of the package. Then, we demonstrate how to use \pkg{gridcp} on a real data example with multiple changepoints.

\subsection{Design and interface}
To describe the design and interface of \pkg{gridcp}, we start with a minimal code example given below that demonstrates the basic usage of the package and the interaction between its main components. The example draws a (synthetic) univariate data stream with a change in the mean after 500 samples, sets up a grid detector (\code{GridDetector}, line 9) using a classical change-in-mean CUSUM score for a single feature (\code{CUSUM}, line 8, see Section~\ref{sec:builtins}), and processes the data one sample at a time until an alarm is raised:



\begin{minted}[linenos, frame=lines]{python}
import numpy as np
from gridcp import GridDetector
from gridcp.scores import CUSUM

signal = np.concatenate([np.repeat(0.0, 500), np.repeat(1.5, 500)])
data = signal + np.random.standard_normal(len(signal))

score = CUSUM(n_features=1)
detector = GridDetector(score=score, threshold=3.0)

state = detector.init_state()
for y in data:
    state, output = detector.update(state, y)
    if output["alarm"]:
        print(f"Alarm at t = {output['n_samples']}")
        break
\end{minted}

Two design features of the \pkg{gridcp} package are worth highlighting.
First, the \code{GridDetector} accepts any score model that implements the \code{ScoreModel} interface, like the built-in \code{CUSUM} score model.
A \code{ScoreModel} encapsulates the running summary statistic $H_t$ and the associated penalized score $S_b^{(t)}/\mathrm{pen}(t)$, while the \code{GridDetector} handles all grid-related bookkeeping.
This separation makes it straightforward to swap in a different score or implement a custom one without any changes to the detector itself.
Section~\ref{sec:builtins} describes all the built-in scores in the package. Details on the \code{ScoreModel} interface and guidance for implementing custom scores can be found in Appendix~\ref{subsec:scoremodel}.

Second, the package employs a \emph{stateless} design.
The detector's only parameters are \code{score} (the score model) and \code{threshold} (the leading constant $\lambda$ in \eqref{overalltest}). 
These are fixed at construction time, and all mutable per-step information is stored in a separate \code{DetectorState} object that is passed to and returned by the \code{GridDetector}'s \code{update} method rather than modified in place.
The main motivation for this design is to facilitate integration with streaming frameworks and distributed systems, such as Apache \pkg{Kafka}, Apache \pkg{Flink}, and Apache \pkg{Spark Streaming}.
In these environments, a mutable state can be difficult to manage and synchronize across multiple workers or concurrent data streams from different sensors, users, or units.
There are, however, several additional practical benefits to the stateless design even outside of distributed settings.
For example, states can easily be serialized and stored in an external database, and restoration, and replay from any timestamp is straightforward.



With this design in mind, the public interface of \code{GridDetector} consists of two methods:
\begin{description}
  \item[\code{init\_state()}:] Initialises the detector's state with an empty grid and a sample count of zero. Called once before the loop (line~11), and can also be used to reset the detector at any point.
  \item[\code{update(state, y)}:] Takes the current state and a new observation \code{y}, updates the grid $B^{(t)}$ and the stored summaries, recomputes the penalized scores over all active split points, checks the threshold, and returns a \code{(new\_state, output)} tuple (line~13).
\end{description}
The \code{output} dictionary contains the following fields:
\begin{itemize}
  \item \code{"n\_samples"}: The number of observations processed so far.
  \item \code{"alarm"}: \code{True} if the detection criterion \eqref{overalltest} is satisfied at the current step, otherwise \code{False}.
  \item \code{"max\_score"}: The maximum penalized score $S_b^{(t)} / \mathrm{pen}(t)$ over the grid $B^{(t)}$.
  \item \code{"max\_split\_point"}: The $0$-indexed integer split point ($b-1$) at which the penalized score was maximized. 
\end{itemize}
To reset a detector after a change has been detected, one may simply call \code{init_state()} to obtain a fresh new detector state, which is demonstrated in Section \ref{sec:well-log}. More complex approaches are discussed briefly in Section~\ref{sec:discussion}.

We remark that both \code{"max\_score"} and \code{"max\_split\_point"} can be either a single number or a vector of length \code{n\_scores}, as a score model is allowed to report more than one value at each split point.
This is useful to allow several statistics to be computed simultaneously, for example one for sparse changes and another for dense changes. The \code{GridDetector} handles this natively.
The number of output values, \code{n\_scores}, is required to be exposed by the score model, and \code{threshold} is then supplied as a vector of the same length, one entry per score. At each step the detector compares every component against its own threshold across all split points, and declares an alarm as soon as at least one of them is exceeded. Therefore, a single-output score is simply the special case \code{n\_scores}~$=1$ with a scalar threshold.

To ensure computational efficiency,
all computationally intensive built-in functions (CUSUM computations, likelihood ratio evaluations, matrix operations) are compiled to machine code via \pkg{Numba}'s \code{@nb.njit} decorator with caching enabled \citep{lam2015numba}.
The first call to a score function incurs a one-time compilation overhead of typically at most a few seconds.
Subsequent calls run at near-native speed, and on a standard laptop, processing 10{,}000 observations of dimension $p=1000$ with the \code{CUSUM} score takes approximately 0.3 seconds after compilation.

\subsection{Example: Detecting multiple changepoints in the well log dataset}\label{sec:well-log}
As an example, we demonstrate the use of \pkg{gridcp} on the univariate well log data from the Turing Change Point Dataset \citep{oruanaidh1996numerical, vandenburg2020evaluation}, which shows nuclear magnetic resonance (NMR) measurements taken at regular depth intervals along a borehole. For simplicity, we assume Gaussian noise with unknown variance, and we wish to detect changes in the mean. Therefore, in lines 1--4 of the code example below, we import and define the \code{GaussianMean} score (defined in  Section~\ref{sec:builtins}), and initialize the corresponding detector. The detection threshold is set to $2.8$, which is the result of a Monte Carlo calibration to a false alarm rate of $5\%$. We explain the available procedures for calibrating the threshold later in Section~\ref{sec:calibration}. 

Next, lines 6--14 involve processing the data. In contrast to the example in the previous section, we now wish to detect several changepoints. To do so, when the detector raises an alarm, we simply reset the detector to its initial state and continue processing new observations. Resetting can be done by running \code{state = detector.init_state()} (line 14), which returns the detector to its original state before any data was observed. Note that this reset happens at the detected \emph{alarm time}, which is at least one observation after the actual changepoint. We discuss the implications of this approach in Section~\ref{sec:discussion}. In the code example, we additionally store the maximum score observed at each time step by extracting \code{"max_score"} from the output object (line 11). 
\begin{minted}[linenos, frame=lines]{python}
from gridcp.scores import GaussianMean

score = GaussianMean()
detector = GridDetector(score = score, threshold = 2.8)

state = detector.init_state()
alarms = []
scores = []
for i, y in enumerate(data):
    state, output = detector.update(state, y)
    scores.append(output["max_score"])
    if output["alarm"]:
        alarms.append(i)
        state = detector.init_state()
\end{minted}

\begin{figure}[ht]
    \centering
    \includegraphics[width=\linewidth]{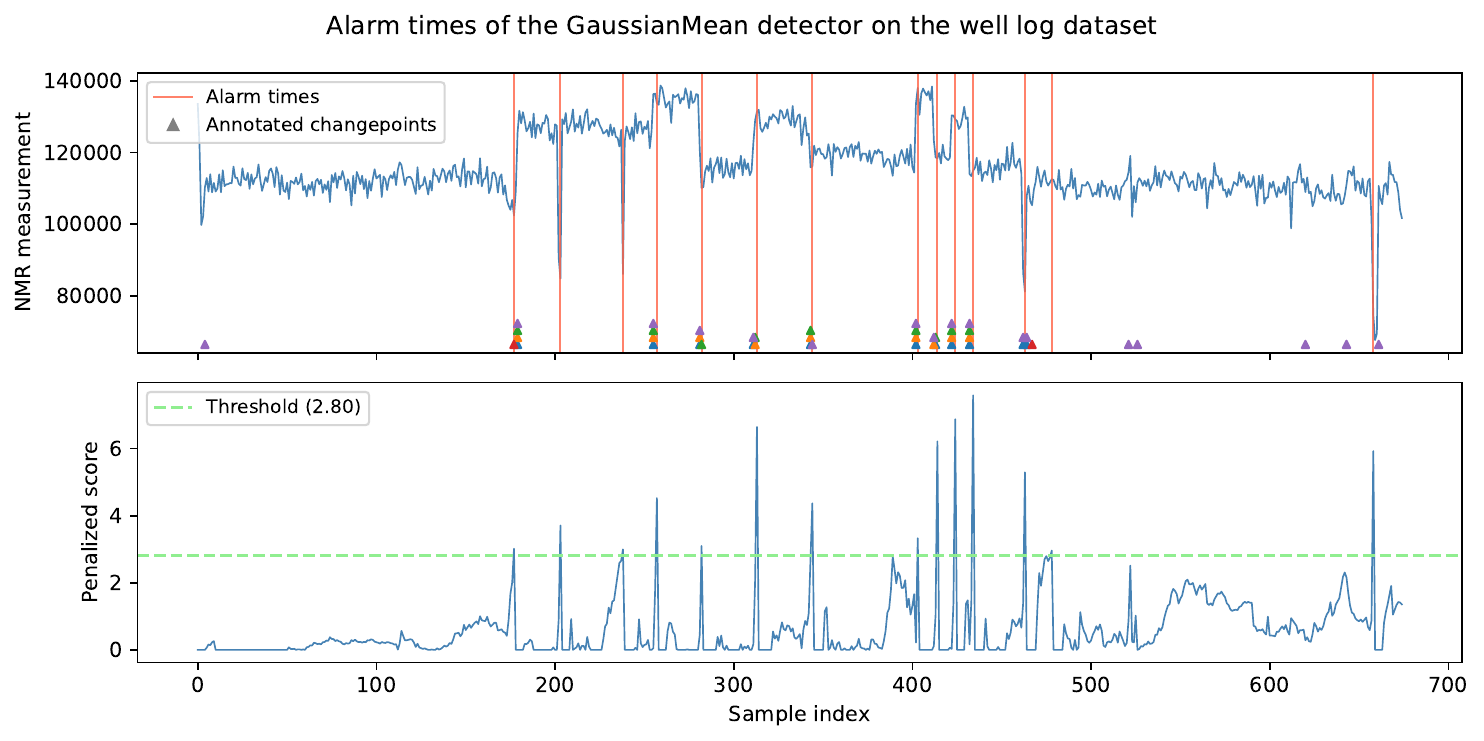}
    \caption{Online changepoint detection on the well log dataset using the \code{GaussianMean} score model with a bootstrap-calibrated threshold. Top: NMR measurements along the borehole, with alarm times (vertical red lines) and annotated changepoints (triangles, one color per annotator). Bottom: Running maximum penalized score, with the calibrated threshold of $2.8$ (dashed green line).}
    \label{fig:well_log}              
\end{figure}

The alarm times for the well log data compared to the annotated changepoints and the maximum score over time compared to the threshold are shown in the top and bottom displays of Figure~\ref{fig:well_log}, respectively. 
One observes that the score resets to zero immediately after exceeding the threshold, which is due to the detector being reset to its initial state upon raising an alarm.


\section{Built-in score models}\label{sec:builtins}

The package \pkg{gridcp} ships with nine built-in score models, covering a broad range of commonly encountered changepoint scenarios. Each implements the \code{ScoreModel} protocol and uses \pkg{Numba}-compiled kernels for inner computational loops. This section gives a high-level overview of the score models, while more details are provided in  Appendix~\ref{app:scoremath}. Table~\ref{tab:builtins} summarizes the  score models, which naturally group into four families: (i) Gaussian score models, (ii) regression score models, (iii) the exponential-family GLR score model, and (iv) the nonparametric NPFOCuS score model. Before we describe each family, we first describe the common structure they share.
\begin{table}[t!]
\centering
\begin{tabular}{l l l}
\hline
Score & Detects change in & Score type \\
\hline
\code{CUSUM}                    & Mean (known variance)          & per-coordinate          \\
\code{GaussianMean}             & Mean (unknown covariance)      & per-coordinate or joint \\
\code{GaussianVariance}         & Variance (zero mean)           & per-coordinate          \\
\code{GaussianMeanOrVariance}   & Mean and/or variance           & per-coordinate          \\
\code{GaussianMeanOrCovariance} & Mean and/or covariance         & joint                   \\
\code{RegressionWald}           & Regression coefficients        & joint                   \\
\code{RegressionMcScan}         & Regression coefficients        & joint                   \\
\code{ExponentialFamilyGLR}     & Natural parameter (exp.\ fam.) & joint                   \\
\code{NPFOCuS}                  & Arbitrary distribution         & per-coordinate          \\
\hline
\end{tabular}
\caption[Summary of built-in score models in \pkg{gridcp}.]{Summary of built-in score models in
\pkg{gridcp}. }
\label{tab:builtins}
\end{table}

\paragraph{Aggregation.}
All built-in score models accept multivariate observations, and aggregate information across the $p$ coordinates in one of two ways. A \emph{per-coordinate} score model computes a
univariate score for each coordinate and combines the $p$ scores through an \code{aggregation} argument. A \emph{joint} score model forms a
single multivariate score directly and takes no such argument. For a per-coordinate score model, no single rule for combining coordinates is best
against every change: a maximum over coordinates is typically powerful when the change is concentrated in a few coordinates but weak when it is spread across many, and a
sum has the opposite profile \citep{tickle,liu_minimax_2021}. The \code{aggregation}
argument therefore offers four options: \code{"max"} returns the coordinate-wise
maximum, targeting sparse changes (default); \code{"sum"} returns the sum, targeting dense
changes; \code{"max-sum"} returns both, so that the detector has power for both sparse and dense changes; and \code{None} returns the $p$ per-coordinate
scores individually. We recall that when a score model outputs multiple score values, the \code{GridDetector} gives each a separate threshold and raises an alarm as soon as any of them exceeds
it (as described in Section~\ref{sec:package}).

\paragraph{Penalty.} 
With the exception of \code{NPFOCuS} and \code{RegressionMcScan}, each built-in score model has a per-split, per-sample-size distribution that is exactly or approximately chi-squared under the null of no change. Each of these therefore admits a natural chi-squared-type penalty of the form \begin{equation}   \mathrm{pen}_{M,\mathrm{df}}(t) = \log(tM) + \sqrt{\mathrm{df}\,\log(tM)}.   \label{penaltylm} \end{equation} Here, $M$ and $\mathrm{df}$ depend on the score model and aggregation (detailed in   Appendix~\ref{app:scoremath}), and are computed automatically. The penalty \eqref{penaltylm} is not a chi-squared quantile, nor in itself a high-probability bound. Rather, it is the \emph{rate} of the chi-squared concentration inequality of \citet{laurentmassart}, when taking the maximum of $M$ chi-squared random variables with $\mathrm{df}$ degrees of freedom. Leading constants are intentionally absorbed into $\lambda$ in \eqref{overalltest}, which can be calibrated by Monte Carlo simulation using the calibration utilities presented in Section~\ref{sec:calibration}. 

We remark that, when the null distribution of the score is exactly chi-squared, a sufficiently large $\lambda$ guarantees false alarm control over an unbounded horizon. This rests on the logarithmic growth of \eqref{penaltylm} in $t$, and is shown in Appendix~\ref{app:scoremath}. 

The time-varying penalty \eqref{penaltylm} can be enabled and disabled through the boolean score model input \code{enable_penalty}, which defaults to \code{True}. It is recommended enabled for false alarm probability calibration (see Section \ref{subsec:calib-fa}), where it makes the detector progressively more conservative so that the false alarm rate stays close to target well beyond the calibration horizon. Setting \code{enable\_penalty = False} replaces it with a constant penalty of $1$, which is appropriate for average run length calibration (Section~\ref{subsec:calib-arl}).
\newpage

\subsection{Gaussian scores}\label{subsec:gaussianscores}

\pkg{gridcp} includes five Gaussian score models, all of which test for a change in the
parameters of the model
\begin{align}
    Y_i \sim \begin{cases}
      \N(\mu_1, \Sigma_1), & i < \tau,\\
      \N(\mu_2, \Sigma_2), & i \geq \tau.
    \end{cases} \label{gaussianmodel}
\end{align}
Each score model implements the
closed-form likelihood-ratio statistic for such a change, and the five differ in
three respects: which parameters may change and what is assumed about the
remaining nuisance parameters (Table~\ref{tab:gaussian}), and whether the score
is per-coordinate or joint (Table~\ref{tab:builtins}). The explicit statistics
and the estimators they use are given in Appendix~\ref{app:gaussmath}.

\begin{table}[h!]
\centering
\resizebox{\textwidth}{!}{
\begin{tabular}{l l l}
\hline
Score model & Parameters of interest & Model assumptions \\
\hline
\code{CUSUM}                    & mean $\mu$           & $\Sigma = I_p$ known \\
\code{GaussianMean}             & mean $\mu$           & $\Sigma$ unknown but constant; diagonal or full \\
\code{GaussianVariance}         & variances           & $\mu = 0$ known; $\Sigma$ diagonal \\
\code{GaussianMeanOrVariance}   & mean and variances  & both unknown; $\Sigma$ diagonal \\
\code{GaussianMeanOrCovariance} & mean and covariance & both unknown; $\Sigma$ full \\
\hline
\end{tabular}}
\caption{The Gaussian score models as likelihood-ratio tests within model
\eqref{gaussianmodel}, listing for each score model the parameters of interest, whose
change is tested, and the assumptions made on the remaining parameters.}
\label{tab:gaussian}
\end{table}

\subsection{Regression score models}\label{subsec:regression}

\pkg{gridcp} includes two regression score models, both of which test for a change in
the coefficient vector of the linear regression model
\begin{align}
    Y_i = \begin{cases}
      X_i^\top \beta_1 + \epsilon_i, & i < \tau,\\
      X_i^\top \beta_2 + \epsilon_i, & i \geq \tau,
    \end{cases} \label{regressionmodel}
\end{align}
with $\beta_1 \neq \beta_2 \in \RR^q$, each observation supplied as a vector
$(Y_i, X_i)$ of the response and the $q$ regressors. Unlike the Gaussian score models,
the two make different assumptions on the covariates $X_i$ and the noise
$\epsilon_i$ (Table~\ref{tab:regression}), and consequently have different null
behavior. \code{RegressionWald} treats the covariates as fixed and the noise as
Gaussian, $\epsilon_i \sim \N(0,1)$. Its two-sample Wald statistic is
\emph{exactly} $\chi^2(q)$ under the null. \code{RegressionMcScan} instead treats
the covariates as random, with zero mean and covariance $\Sigma$, and the noise
as sub-Gaussian. The resulting ``McScan'' statistic of \citet{cho2025detection}
is a maximum over the $q$ regressors, has no closed-form null distribution, and is
penalized by $\sqrt{\log(qt)}$, a high-probability bound taken from the same
paper. Both score models are joint, and do not accept any \code{aggregation} argument. Their explicit statistics are given in Appendix~\ref{app:regmath}.

\begin{table}[t!]
\centering
\begin{tabular}{l l l}
\hline
Score model & Covariates & Noise \\
\hline
\code{RegressionWald}   & fixed  & Gaussian, $\epsilon_i \sim N(0,1)$ \\
\code{RegressionMcScan} & random & sub-Gaussian \\
\hline
\end{tabular}
\caption{The two regression score models for a change in the coefficient vector $\beta$
of the linear model \eqref{regressionmodel}, and their differing assumptions on
the covariates and the noise.}
\label{tab:regression}
\end{table}

\subsection[ExponentialFamilyGLR]{The \code{ExponentialFamilyGLR} score model}\label{subsec:expfamglr}
The score model \code{ExponentialFamilyGLR} tests for a change in the natural parameter $\theta$ of a canonical exponential family,
\begin{align}
    Y_i \sim \begin{cases}
      f(\,\cdot \mid \theta_1), & i < \tau,\\
      f(\,\cdot \mid \theta_2), & i \geq \tau,
    \end{cases} \label{expfammodel}
\end{align}
with $\theta_1 \neq \theta_2$, where
\begin{equation}\label{eq:expfam}
    f(y \mid \theta) = c(y)\exp\bigl(\theta^\top h(y) - A(\theta)\bigr)
\end{equation}
is the exponential-family density, $h$ the sufficient statistic, $v$ its dimension, $A$ the log-partition function, and $c$ the base measure. The outputted score is the centered generalized log-likelihood-ratio statistic for such a change, which is approximately $\chi^2(v)$ under the null. The score model is joint: it forms a single multivariate statistic across all coordinates and therefore accepts no \code{aggregation} argument. Several of the Gaussian score models are closed-form special cases, whilst \code{ExponentialFamilyGLR} instead covers the general case at the cost of solving a small system of equations by Newton's method at each split point, and should therefore be used for families the other score models do not cover. Unlike the other built-in score models, it requires the user to supply the exponential family, either by name through the \code{from\_family} class method, drawing from the built-in families in Table~\ref{tab:expfam-families}, or by providing $h$, $A$, and the derivatives of $A$ directly. The explicit statistic and estimators, together with details on specifying a custom family, are given in Appendix~\ref{app:glrmath}.

\begin{table}[t!]
\centering
\resizebox{\textwidth}{!}{
\begin{tabular}{lll}
\hline
Family & Detects change in & Additional parameters \\
\hline
\code{gaussian\_mean}           & Gaussian mean (known variance $=1$)    & \code{n\_features} \\
\code{gaussian\_variance}       & Gaussian variance (known mean $=0$)    & --- \\
\code{gaussian\_mean\_variance} & Gaussian mean and variance             & --- \\
\code{gaussian\_covariance}     & Gaussian covariance (known mean $=0$)  & \code{n\_features} \\
\code{poisson}                  & Poisson rate                           & --- \\
\code{exponential}              & Exponential rate                       & --- \\
\code{bernoulli}                & Bernoulli success probability          & --- \\
\code{gamma\_rate}              & Gamma rate (known shape)               & \code{shape} \\
\hline
\end{tabular}
}
\caption{Built-in families available with \code{ExponentialFamilyGLR.from\_family}. For families that support multivariate data, \code{n\_features} (default $1$) sets the dimension; for the remaining families \code{n\_features = 1} is the only accepted value. The Gamma \code{shape} parameter defaults to $1.0$.}
\label{tab:expfam-families}
\end{table}

\subsection[NPFOCuS]{The \code{NPFOCuS} score model}\label{subsec:npfocus}
\code{NPFOCuS} tests for an arbitrary change in the distribution of the data,
\begin{align}
    Y_i \sim \begin{cases}
      P_1, & i < \tau,\\
      P_2, & i \geq \tau,
    \end{cases} \label{npfocusmodel}
\end{align}
with $P_1 \neq P_2$ and no parametric form assumed. It discretizes each coordinate against a fixed, user-supplied ``value grid'' $\mathcal U=\{u_1,\dots,u_m\}$ (not to be confused with the grid $\mathcal{B}$). For each element of $\mathcal U$, the \code{NPFOCuS} score computes a likelihood-ratio statistic for a change in the probability that an observation falls below it, implementing the per-split statistic of \citet{npfocus}. Within each coordinate, these statistics are combined over the values in $\mathcal U$ by both summing and maximizing, and the two resulting statistics are then aggregated across coordinates according to the \code{aggregation} argument.  \code{NPFOCuS} is therefore a per-coordinate score. Unlike the chi-squared scores, \code{NPFOCuS} applies no penalty, as its statistics admit no natural time-dependent penalty. The likelihood-ratio statistics are given in Appendix~\ref{app:npfocusmath}.

\subsection[GRB-detection]{Illustration of \code{ExponentialFamilyGLR} and \code{NPFOCuS}} 

Our earlier data example (Section~\ref{sec:well-log}) considered a well-studied change-in-mean problem. To illustrate some of the other score models detailed above, we now apply two of these in a count-data application borrowed from \citet{ward2025poissonfocus}: the detection of gamma-ray bursts. The FERMI gamma-ray burst data consists of gamma-ray photon counts coming from sensors on the NASA Fermi Gamma-ray Space Telescope, and the task is to detect unusual increases in the photon count relative to the baseline count coming from background radiation. As this is count data, a Poisson model is a natural choice, and a gamma-ray burst then corresponds to an increase in the Poisson rate \citep[see][for details]{ward2025poissonfocus}. The full dataset contains recordings of several bursts, and for a simple illustration we focus on a single one, namely GRB171004857. The rest of the time series are stored in the variable \code{training_data}, and we use these to estimate the background radiation mean and standard deviation and to calibrate the detectors to a desired false alarm rate. Each time series consists of 6 sensors which we sum over to obtain univariate series. 

For this data set, a natural choice of score model is \code{ExponentialFamilyGLR} specified with a Poisson distribution. A nonparametric alternative is \code{NPFOCuS}, which makes no distributional assumption at all. In this example, we employ both, as shown in the code chunk below. To define the \code{NPFOCuS} detector, one needs to specify $\mathcal{U}$ (see Section \ref{subsec:npfocus}). Our approach is to estimate the mean background photon count from historical data (\code{training_data}) using a robust estimator and then call \code{np.linspace} to make a symmetric grid around this value. Below, we use a robust estimator for the background standard deviation ($1.4826\cdot \mathrm{MAD}$, where $\mathrm{MAD}$ denotes the median absolute deviation), to determine a reasonable width of $\mathcal{U}$. For the Poisson detector, one needs to specify an initial value of the unknown rate parameter (\code{theta_init}) when constructing the \code{ExponentialFamilyGLR} score. Since the Poisson natural parameter is $\theta = \log\mu$, we initialize it as the logarithm of the estimated background rate: 

\begin{minted}[linenos, frame=lines]{python}
from gridcp.scores import ExponentialFamilyGLR, NPFOCuS

robust_lambda = float(np.median(training_data))
robust_scale = float(
    1.4826 * np.median(np.abs(training_data - robust_lambda))
)

poisson_theta_init = np.log(robust_lambda)
glr_score = ExponentialFamilyGLR.from_family(
    "poisson",
    n_features = 1,
    theta_init = poisson_theta_init,
    enable_penalty = False,
)
value_grid = np.linspace(
    robust_lambda - robust_scale * 2,
    robust_lambda + robust_scale * 2,
)
npf_score = NPFOCuS(value_grid = value_grid, n_features = 1)
\end{minted}

\begin{figure}
    \centering
    \includegraphics[width=1\linewidth]{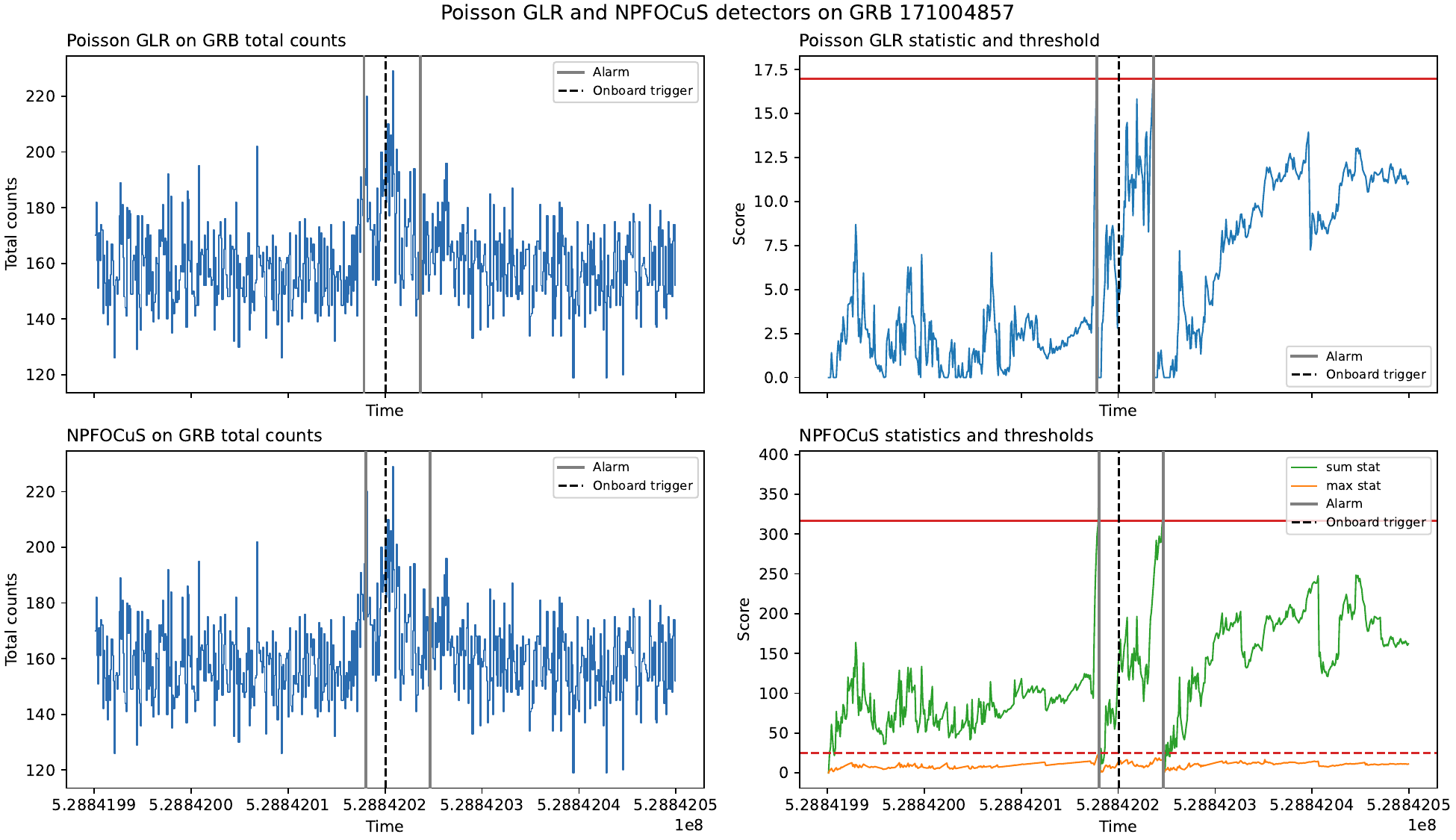}
    \caption{Left column: The alarm time for a Poisson change-in-rate detector from \code{ExponentialFamilyGLR} and the nonparametric detector \code{NPFOCuS} when applied to a known gamma-ray burst. The black dashed line is the onboard trigger of the Fermi telescope. Right column: The detector statistics and their associated thresholds.}
    \label{fig:FERMIDETECTION}
\end{figure}

Next, we need to define the thresholds to be used by the detectors, and moreover the detector objects themselves. In this example, the detectors were calibrated to a false alarm probability of 5\% using a Poisson sampler with rate \code{robust_lambda} (see Section~\ref{sec:calibration} for details on calibration), omitted here for brevity:
\begin{minted}[linenos, frame=lines]{python}
glr_detector = GridDetector(
    score = glr_score, threshold = 16.96
)
npf_detector = GridDetector(
    score = npf_score, threshold = np.array([317.07, 24.88])
)
\end{minted}
Finally, we can run the detectors over the GRB171004857 series (stored as \code{grb_total}) using a \code{for}-loop, keeping track of potential alarms as we go:
\begin{minted}[linenos, frame=lines]{python}
glr_state = glr_detector.init_state()
glr_alarms = []
npf_state = npf_detector.init_state()
npf_alarms = []
for i, y in enumerate(grb_total):
    glr_state, glr_out = glr_detector.update(glr_state, y)
    if glr_out["alarm"]:
        glr_alarms.append(i)
        glr_state = glr_detector.init_state()

    npf_state, npf_out = npf_detector.update(npf_state, y)
    if npf_out["alarm"]:
        npf_alarms.append(i)
        npf_state = npf_detector.init_state()
\end{minted}
As we can see from the output in Figure~\ref{fig:FERMIDETECTION}, both methods successfully detect the gamma-ray burst, with the Poisson detector being slightly quicker. Both detectors trigger significantly earlier than NASA's onboard algorithm~\citep[dashed line, see, e.g.,][]{Meegan2009FermiGBM} and quickly detect the return to background radiation. 

\section{Simulation and threshold calibration}\label{sec:calibration}
A critical step before deploying any online changepoint detector is choosing the detection threshold $\lambda$ in \eqref{overalltest} to achieve a desired level of false alarm control or average run length. \pkg{gridcp} provides Monte Carlo-based methods for calibrating the threshold against two different targets: the false alarm probability over a finite horizon (Section~\ref{subsec:calib-fa}) and the average run length under the null (Section~\ref{subsec:calib-arl}). Each target can be calibrated either from a user-supplied sampler of the null distribution, or directly from data when only a change-free data stream or a pre-computed set of null sample paths is available. All of these Monte Carlo simulations are parallelized. We first describe both calibration targets under the assumption of a known null distribution, and then show in Section~\ref{sec:realdata} how the same methods can be applied to calibrate from data. In Section~\ref{subsec:simulation-highd} we brings these tools together in a high-dimensional simulated example.


\subsection{False alarm probability calibration}\label{subsec:calib-fa}
Recall that the false alarm probability of an online changepoint detector is the probability that it raises an alarm on a null stream of infinite length, i.e., $\PP_{\infty}(\tauhat <\infty)$. In practice, however, precise false alarm control can only be achieved via Monte Carlo simulation over finite sequences. Given a maximum sequence length $T$, a realistic goal is to instead target $\PP_{\infty}(\tauhat\leq T)$.

Let $\delta \in (0,1)$ denote a desired false alarm probability. In \pkg{gridcp}, the utility function \code{calibrate\_threshold\_false\_alarm} uses Monte Carlo samples to calibrate the threshold $\lambda$ to achieve the desired level $\delta$. Specifically, this function returns the empirical $(1 - \delta)$-quantile of the path-wise maximum penalized score distribution
\begin{align}
    \max_{t = 2,3,\ldots, T} \ \max_{b \in B^{(t)}} \ \frac{S_b^{(t)}}{\mathrm{pen}(t)}
\end{align} 
by simulating \code{n\_paths} null paths from a user-supplied sampler, where $S_b^{(t)}$ and $\mathrm{pen}(t)$ respectively denote the score and penalty for a given score model. By default, score models are defined with the time-dependent penalty $\mathrm{pen}(t)$, set through the parameter \code{enable\_penalty = True}. The contrasting setting \code{enable\_penalty = False} is used for average run length calibration, and is described in Section~\ref{subsec:calib-arl}. 

Below is an example where we calibrate the threshold of the CUSUM statistic using \code{n\_paths} = $20{,}000$ standard Gaussian sequences to achieve a false alarm probability of \newline $\delta = $ \code{false\_alarm\_probability} $=0.05$ for $T=$ \code{stream\_len} $= 100$:
\begin{minted}[linenos, frame=lines]{python}
from gridcp.calibration import calibrate_threshold_false_alarm

score = CUSUM(n_features = 1, enable_penalty = True)
threshold = calibrate_threshold_false_alarm(
    score,
    false_alarm_probability = 0.05,
    n_paths = 20_000,
    stream_len = 100,
    pre_sampler = lambda rng: rng.standard_normal(),
    rng = 0,
)
detector = GridDetector(score = score, threshold = threshold)
\end{minted}
Here, the null distribution is specified via the \code{pre\_sampler} argument, a callable that draws a single observation and takes a \code{numpy.random.Generator} as input. The \code{rng} argument on line 10, distinct from the generator handed to \code{pre\_sampler}, makes the calibration reproducible, and accepts either a \code{numpy.random.Generator} or an integer seed. Once the threshold has been estimated, the detector is constructed as usual (line 12). 

The null paths are distributed across all available CPU cores by default. The number of worker processes can be set through the \code{n\_jobs} argument, and parallelism can be disabled altogether with \code{parallel = False}. With a fixed \code{rng}, results from parallel execution are only exactly reproducible provided \code{n\_jobs} is unchanged, since the paths are split across workers with distinctly derived rngs. However, passing \code{strict\_equivalence = True} guarantees identical results for any value of \code{n\_jobs}, although at a modest cost in speed.

For score models that output multiple scores, a Bonferroni correction is applied by default, setting each per-component threshold at the $(1 - \delta/$\code{n\_scores}$)$-quantile. This can be disabled with \code{apply\_bonferroni = False}.

In general, the threshold generated by \code{calibrate\_threshold\_false\_alarm} can only be guaranteed to target the desired false alarm probability over the given sequence length \code{stream\_len}. However, for certain time-varying penalties, the false alarm guarantee may hold approximately even for sequences longer than \code{stream\_len}. For example, for the univariate ($p=1$) CUSUM statistic implemented in \code{CUSUM}, it can be shown that $\PP_{\infty}(\tauhat < \infty) \leq \delta$ whenever $\mathrm{pen}(t)=\log t + \sqrt{\log t}$ and the threshold $\lambda$ is chosen sufficiently large \citep{anote,moen_grid}. Since the penalty $\mathrm{pen}(t)$ is increasing in $t$, the detector grows more conservative as $t$ increases: thus, $\PP_{\infty}(\tauhat=t)$ decreases with $t$. Heuristically, this means that, if \code{stream\_len} is chosen adequately large, we may achieve $\PP_{\infty}(\tauhat < N) \approx \PP(\tauhat < $ \code{stream\_len} $)$ even when $N\gg $ \code{stream\_len}. 

\begin{figure}[ht]                                    \centering
    \includegraphics[width=\linewidth]{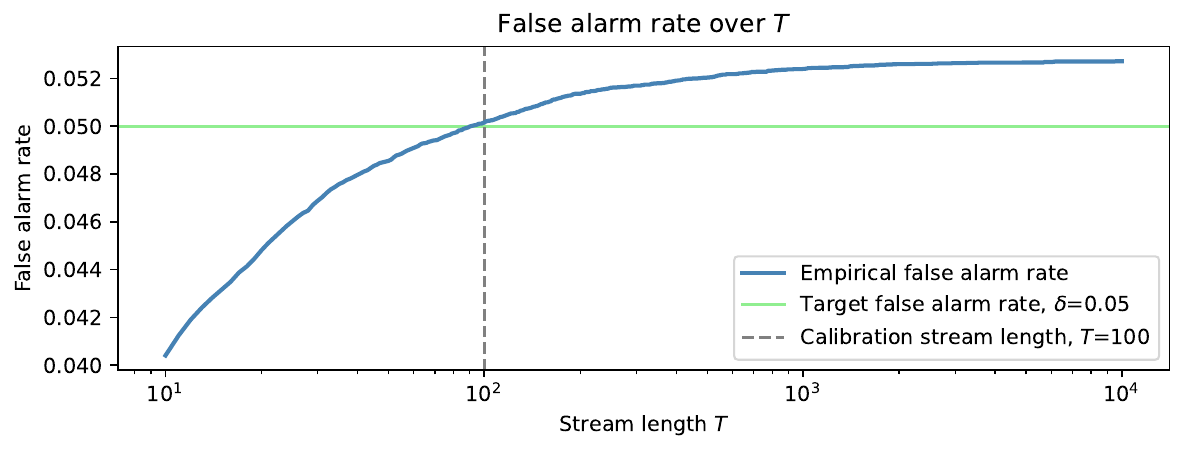}
    \caption{Empirical false alarm rate of the \code{CUSUM} score as a function of stream length $T$, across $100{,}000$ null streams. The threshold is calibrated to a target false alarm rate of $\delta = 0.05$ (green line) at a calibration stream length of $T = 100$ (dashed gray line).}
    \label{fig:fa_rate}
\end{figure}

To illustrate this behavior, we run the \code{CUSUM} detector that we calibrated earlier over $100{,}000$ paths of length $10{,}000$ from the null distribution, that is, streams a hundred times longer than what the detector is calibrated on. We do this using one of \code{gridcp}'s lower-level simulation utilities, \code{mc\_alarm\_times}, which returns each path's first alarm time. We remark that two other utility functions are available for similar use cases: \code{draw\_samples} for generating streams and \code{mc\_max\_scores} for generating path-wise maximum scores. Once complete, we aggregate the alarm times into false alarm rates over a geometric grid of $5{,}000$ time points from $10$ to $10{,}000$ (lines 10--12).
\begin{minted}[linenos, frame=lines]{python}
from gridcp.calibration import mc_alarm_times

alarm_times = mc_alarm_times(
    detector, 
    n_paths = 100_000, 
    stream_len = 10_000, 
    pre_sampler = lambda rng: rng.standard_normal(), 
    rng = 1, 
)
Ts = np.geomspace(10, 10_000, 5_000).astype(int)
at_sorted = np.sort(alarm_times)
false_alarm_rates = np.searchsorted(at_sorted, Ts) / 100_000
\end{minted}
The empirical false alarm rate over time is shown in Figure~\ref{fig:fa_rate}. Here, we observe that it remains close to the target even for streams substantially longer than the calibration stream length, confirming the conservative behavior of the time-dependent penalty.


\subsection{Average run length calibration}\label{subsec:calib-arl}
Recall that the average run length of an online changepoint detector $\tauhat$ is given by $\EE_{\infty}(\tauhat)$, i.e., the expected number of observations in the no-change regime until the detector raises an alarm.  
Unlike the false alarm probability, obtaining control over the average run length does not require a time-varying penalty $\mathrm{pen}(t)$, which typically serves to make the detector become more conservative over time. On the contrary, a constant penalty $\mathrm{pen}(t)= 1$ allows us to approximate the time to a false alarm by an exponential distribution, as argued by \cite{chen_high-dimensional_2022}. 

Let $\mathrm{ARL}_0 \in \NN$ denote a desired average run length. In \pkg{gridcp}, the utility function \code{calibrate\_\allowbreak threshold\_arl} uses the Monte Carlo approach of \cite{chen_high-dimensional_2022} to calibrate the threshold $\lambda$ to approximately achieve the desired $\mathrm{ARL}_0$. Specifically, for single-output scores, the function returns the empirical $(1/e)$-quantile of the path-wise maximum score distribution
\begin{align}
    M_{\mathrm{ARL}_0}\;=\;\max_{t = 2,3,\ldots, \mathrm{ARL}_0} \ \max_{b \in B^{(t)}} \ S_b^{(t)}
\end{align}
by simulating \code{n\_paths} null paths of length $\text{ARL}_0$ from a user-supplied sampler. Under the exponential approximation discussed above, a threshold $\lambda$ satisfying $\PP_{\infty}(M_{\mathrm{ARL}_0} \leq \lambda) = 1/e$
yields an expected alarm time of approximately $\mathrm{ARL}_0$. This approach requires the score statistic to be stationary under the null, which in \pkg{gridcp} means setting \code{enable\_penalty = False} on the score model. A \code{UserWarning} is issued if any other penalty type is detected. Below is an example where we calibrate the threshold of the CUSUM statistic under $20{,}000$ standard Gaussian sequences to target $\mathrm{ARL}_0 = $ \code{target\_arl} $=100$:
\begin{minted}[linenos, frame=lines]{python}
from gridcp import calibrate_threshold_arl

score = CUSUM(n_features = 1, enable_penalty = False)
threshold = calibrate_threshold_arl(
    score,
    target_arl = 100,
    n_paths = 20_000,
    pre_sampler = lambda rng: rng.standard_normal(),
    rng = 0,
)
\end{minted}
For score models that output multiple scores ($K>1$), a two-step procedure is used. First, per-test $(1/e)$-quantile thresholds $\widehat{\lambda}_1, \ldots, \widehat{\lambda}_K$ are estimated from the marginal max-score distributions, as in the single score case. Second, the scores are standardized by these per-test thresholds and a common scaling factor $c$ is computed as the $(1/e)$-quantile of the path-wise maxima of the standardized scores. The final thresholds are $c \cdot \widehat{\lambda}_k$ for each test $k=1,\ldots,K$. This second step ensures that the joint average run length, rather than each marginal one, approximately equals $\mathrm{ARL}_0$ \citep[for further details, see][]{chen_high-dimensional_2022}. By default, the same null paths are used for both steps. Setting \code{resimulate\_combined\_threshold = True} uses independent paths instead, removing the potential bias from reusing samples at the cost of twice the computation. In practice, this bias is typically small, hence the default is \code{False}. Parallelism is controlled through \code{parallel} and \code{n\_jobs}, as in Section~\ref{subsec:calib-fa}.

\begin{figure}[ht]
    \centering
    \includegraphics[width=\linewidth]{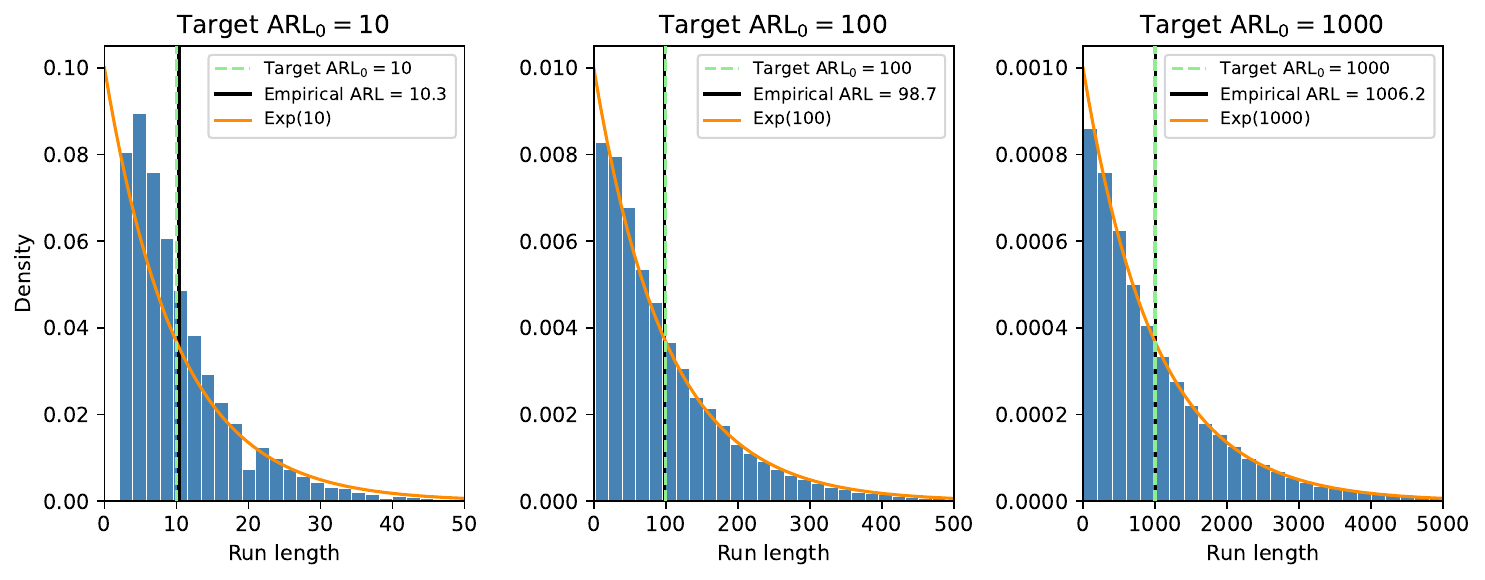}
    \caption{Run length distributions under the null for \code{CUSUM} with \code{enable\_penalty = False} and target $\mathrm{ARL}_0 \in \{10, 100, 1{,}000\}$. For each target, the threshold is calibrated using $20{,}000$ null streams of length $\mathrm{ARL}_0$, and the run lengths are recorded on a separate set of $100{,}000$ null streams of length $10 \cdot \mathrm{ARL}_0$. The orange curves show the corresponding exponential densities.}
    \label{fig:arl_sim}
\end{figure}

While the calibration approach implemented in \code{calibrate\_threshold\_arl} relies on an approximation \citep[see][]{chen_high-dimensional_2022}, the resulting run length distribution aligns closely with its exponential approximation in practice. Figure~\ref{fig:arl_sim} shows the distribution of run lengths under the null for three \code{CUSUM} detectors with target ARLs set to $10$, $100$ and $1{,}000$, respectively. For each \code{CUSUM} detector, the threshold is calibrated using $20{,}000$ null streams of length $\mathrm{ARL}_0$, similar to what was done in the previous code example. Run lengths are then recorded on a separate set of $100{,}000$ null streams of length $10 \cdot \mathrm{ARL}_0$ using \code{mc_alarm_times} (similar to the previous section). The resulting distributions are shown in Figure~\ref{fig:arl_sim}. We observe that the quality of the exponential approximation improves with increasing $\mathrm{ARL}_0$: at $\mathrm{ARL}_0 = 10$ the histogram visibly deviates from the exponential density, while for the two larger targets no significant deviation is visible. Nonetheless, the empirical ARLs align closely with each respective target.


\subsection{Data-driven calibration}\label{sec:realdata}
When the null distribution of the data is not known, or a reasonable parametric sampler is unavailable, \pkg{gridcp} provides two data-driven alternatives: one that takes as input a change-free data stream, and one that takes as input an array of null sample paths. Both follow the same calibration logic as their sampler-based counterparts in Sections~\ref{subsec:calib-fa} and~\ref{subsec:calib-arl}. The difference between them is in how the null samples are generated. Here too, \code{parallel} and \code{n\_jobs} govern the parallel evaluation.

\paragraph{Calibrating from a change-free data stream.} When a change-free data stream is available, the \code{calibrate\_threshold\_false\_alarm\_from\_data} and \code{calibrate\_threshold\_arl\_\allowbreak from\_data} functions generate null paths using circular block bootstrapping \citep{politis1992circular} of the change-free data. Bootstrap streams are constructed by concatenating randomly selected overlapping blocks from the training data, wrapping circularly at the array boundary, which preserves short-range temporal dependence. The block length controls the trade-off between preserving dependence and bootstrap variability: \code{block\_length = 1} gives an i.i.d.\ bootstrap, while \code{block\_length = None} (the default) selects $\max(1, \lfloor T^{1/3} \rfloor)$ automatically, where $T$ is the length of the training data, following the asymptotically optimal rate under weak dependence \citep{lahiri2003resampling}. Below is an example where we calibrate the threshold by targetting the false alarm probability, using a change-free training array \code{training\_data}:
\begin{minted}[linenos, frame=lines]{python}
from gridcp import calibrate_threshold_false_alarm_from_data
threshold = calibrate_threshold_false_alarm_from_data(
    score,
    training_data = training_data,
    false_alarm_probability = 0.05,
    stream_len = 500,
    n_paths = 2000,
    rng = 0,
)
\end{minted}
The ARL counterpart \code{calibrate\_threshold\_arl\_from\_data} works in the same way, taking \code{target\_arl} in place of \code{false\_alarm\_probability} and \code{stream\_len}.

\paragraph{Calibrating from null samples.} When null sample paths have already been generated externally, \code{calibrate\_threshold\_false\_alarm\_from\_samples} and \code{calibrate\_threshold\_\allowbreak arl\_from\_samples} accept an array of null paths of shape (\code{n\_paths} $\times$ \code{stream\_len} $\times$ \allowbreak\code{n\_features}) directly, and use the provided null samples as-is. This gives the user full control over which null paths are used for calibrating the threshold.

\subsection{Simulated example: High-dimensional Gaussian data}\label{subsec:simulation-highd}

We close this section with a larger simulated example that uses \pkg{gridcp}'s calibration and simulation tools in a high-dimensional change-in-mean setting, using the built-in \code{CUSUM} score model with the \code{"max-sum"} aggregation (Section~\ref{subsec:gaussianscores}). The example illustrates how quickly the detector is able to detect a change, as a function of the change magnitude and sparsity, when calibrated to either a target false alarm probability or average run length.  We consider $p = 1{,}000$-dimensional Gaussian data $Y_i \sim \mathrm{N}(\mu_i, I_p)$, with a stream length of $T = 1{,}000$ and a changepoint at $\tau = 500$. The pre-change mean is the null vector, while the post-change mean $\mu$ has $\ell_2$-norm $\normm{\mu}_2$ concentrated equally across the first $s$ coordinates, with sparsity levels $s \in \{1, 10, 1000\}$. The case $s = 1$ corresponds to a fully sparse shift, while $s = p = 1{,}000$ corresponds to a fully dense shift.

We first fix the problem constants and define a Gaussian sampler, following the interface described in Section~\ref{subsec:calib-fa}, where the mean is supplied as an argument to the sampler:

\begin{minted}[linenos, frame=lines]{python}
P, N, TAU = 1000, 1000, 500
SPARSITY = [1, 10, P]
PHI = np.linspace(0.0, 4.0, 17)
NULL_MEAN = np.zeros(P)

def gaussian_sampler(rng, mean):
    return rng.normal(loc = mean, scale = 1.0, size = len(mean))
\end{minted}

We compare two detectors based on the \code{CUSUM} score model: one calibrated to a false alarm probability of $\delta = 0.05$ by setting \code{enable\_penalty = True}, the other to an average run length of $\mathrm{ARL}_0 = 1{,}000$ by setting \code{enable\_penalty = False}. Both thresholds are estimated from $1{,}000$ null Monte Carlo paths, using the calibration functions of Sections~\ref{subsec:calib-fa} and~\ref{subsec:calib-arl}:
\begin{minted}[linenos, frame=lines]{python}
score_faprob = CUSUM(
    n_features = P, aggregation = "max-sum", enable_penalty = True
)
threshold_faprob = calibrate_threshold_false_alarm(
    score_faprob, false_alarm_probability = 0.05,
    n_paths = 1000, stream_len = N,
    pre_sampler = gaussian_sampler, pre_kwargs = {"mean": NULL_MEAN},
)
detector_faprob = GridDetector(
    score = score_faprob, threshold = threshold_faprob
)

score_arl = CUSUM(
    n_features = P, aggregation = "max-sum", enable_penalty = False
)
threshold_arl = calibrate_threshold_arl(
    score_arl, target_arl = N, n_paths = 1000,
    pre_sampler = gaussian_sampler, pre_kwargs = {"mean": NULL_MEAN},
)
detector_arl = GridDetector(
    score = score_arl, threshold = threshold_arl
)
\end{minted}

The detection delay is then estimated over the grid of sparsities and signal strengths. For each pair $(s, \phi)$, we construct the corresponding post-change mean $\mu$ and pass it to \code{mc\_alarm\_times} through \code{post\_kwargs}. The \code{changepoint} argument places $\tau$ within each simulated path, and streams that never alarm return \code{stream\_len}, so their detection delay is capped at the maximal value $T - \tau = 500$:

\begin{minted}[linenos, frame=lines]{python}
delay = {}
detectors = {"FaProb": detector_faprob, "ARL": detector_arl}
for name, detector in detectors.items():
    for s in SPARSITY:
        for phi in PHI:
            mu = np.zeros(P)
            mu[:s] = phi / np.sqrt(s)
            alarm_times = mc_alarm_times(
                detector, n_paths = 1000, stream_len = N, changepoint = TAU,
                pre_sampler = gaussian_sampler,
                pre_kwargs = {"mean": NULL_MEAN},
                post_sampler = gaussian_sampler,
                post_kwargs = {"mean": mu}, 
            )
            d = alarm_times - TAU
            delay[name, s, phi] = float(np.mean(d[d >= 0]))
\end{minted}

\begin{figure}[h!]
    \centering
    \includegraphics[width=\linewidth]{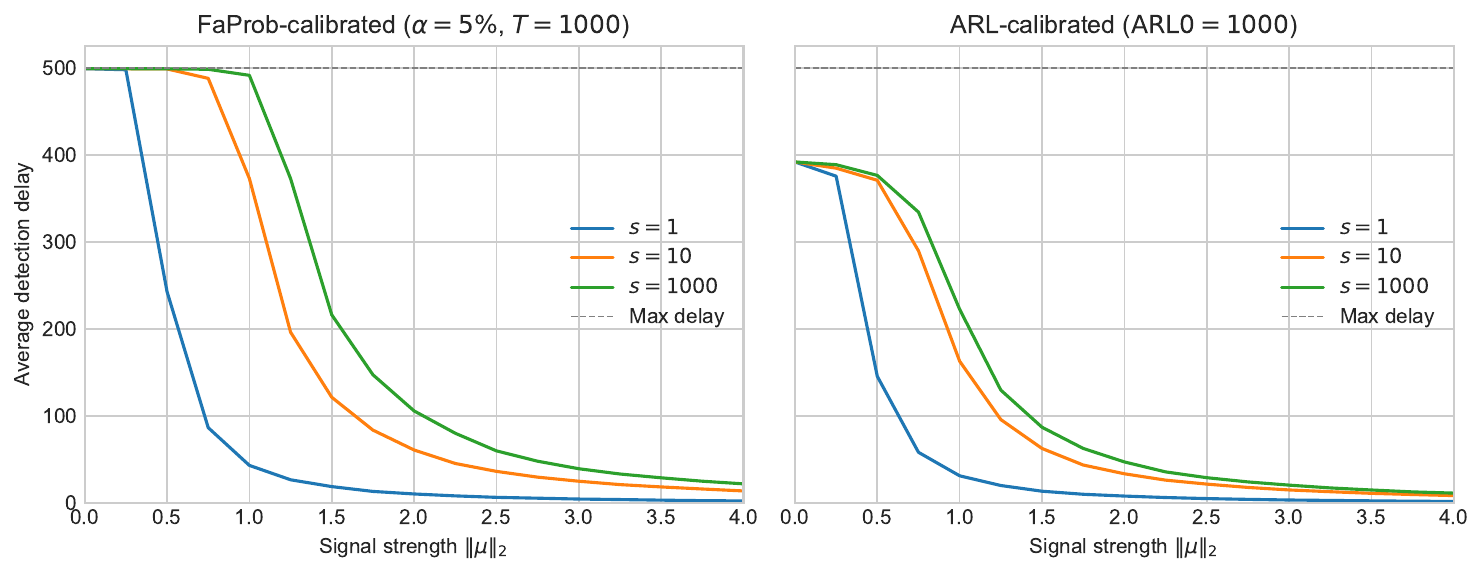}
    \caption{Mean detection delay vs.\ signal strength $\normm{\mu}_2$ for $p = 1{,}000$-dimensional Gaussian data with changepoint at $\tau = 500$. Left:  Detector calibrated to a false alarm probability of $5\%$. Right: Detector calibrated to an average run length of $1{,}000$. Each panel shows three sparsity levels $s \in \{1, 10, 1000\}$. The dashed line marks the maximum possible delay $T - \tau = 500$.}
    \label{fig:highd-delay}
\end{figure}

Figure~\ref{fig:highd-delay} shows the resulting average detection delay as the signal strength $\phi = \normm{\mu}_2$ ranges over $\{0, 0.25, \ldots, 4\}$. Not surprisingly, sparse changes ($s = 1$) are detectable at smaller signal strengths than dense ones. While the two detectors yield broadly consistent curves, the detector calibrated to average run length attains a lower detection delay than the one calibrated to false alarm probability, which is primarily due to the latter being more conservative by construction.

\begin{figure}[ht]
    \centering
    \includegraphics[width=0.55\linewidth]{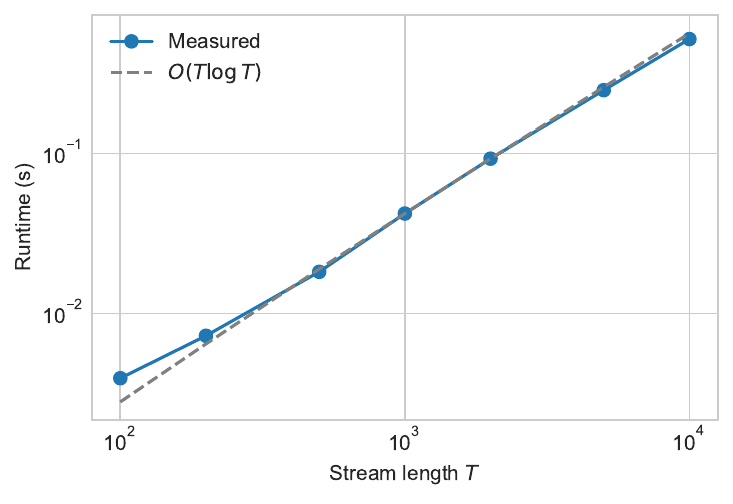}
    \caption{Runtime (seconds) vs.\ stream length $T$ on log-log axes for \code{CUSUM} with $p = 1{,}000$ for the detector calibrated to false alarm probability. The dashed line is an $\mathcal{O}(T \log T)$ reference curve anchored at $T= 1{,}000$.}
    \label{fig:highd-runtime}
\end{figure}

Finally, we measured how the total run time scales with the stream length $T$. Feeding observations to the false-alarm-calibrated detector one at a time, we timed it over a range of stream lengths from $T = 100$ to $T = 10{,}000$, with the result shown in Figure~\ref{fig:highd-runtime}. The total processing time scales approximately as $\mathcal{O}(T\log T)$ in the stream length, consistent with the logarithmic per-observation update times established in \cite{moen_grid}. As a concrete illustration of the computational efficiency of \pkg{gridcp}, we note that processing $10{,}000$ observations in $p = 1{,}000$ dimensions took approximately $0.04$ms per observation ($0.4$ seconds in total) on a single core on a Macbook Pro with an Apple M5 Pro CPU. 

\section{Data example: Sound level readings}\label{sec:soundsending}
As a final data example, we apply \pkg{gridcp} to large-scale industrial vibration data from the Norwegian company Soundsensing AS. 
The dataset consists of Leq (equivalent continuous sound level) decibel measurements from sensors located in different technical rooms housing heating and ventilation systems (HVAC) in commercial buildings. There are approximately 2 years of per-minute data from each sensor, giving a total of around 1 million readings for each sensor. The purpose of changepoint detection in this application is to detect state transitions quickly, for example to indicate that the HVAC system has turned off or on.

\begin{figure}[t]
    \centering
    \includegraphics[width=\linewidth]{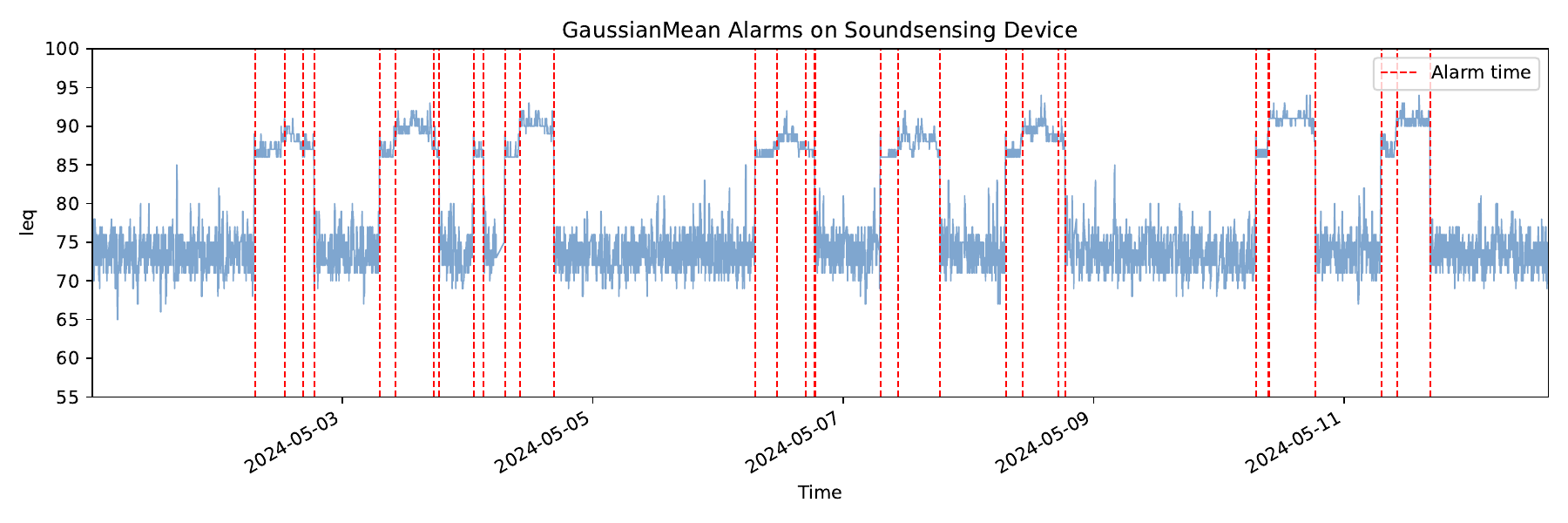}




    \caption{Alarms raised by \code{GaussianMean} for one of the HVAC sound level sensor series. The detector is calibrated to a false alarm rate of 0.001 using the circular block bootstrap method from Section~\ref{sec:calibration}, and the detector is reset immediately after each alarm.}
    \label{fig:Soundsensing_alarm_plots}
\end{figure}

The result of running the \code{GaussianMean} detector on one of the sound level sensor series is illustrated in Figure \ref{fig:Soundsensing_alarm_plots}.
We see that the detector immediately catches all the transitions between high (around 90 dB) to low (around 70 dB) sound levels, corresponding to on and off states, respectively.
The detector is also sensitive to local changes within each high sound level region, but these can be filtered out based on the size of the changes, for example. 
The circular block bootstrap method with default block size from Section \ref{sec:calibration} was used for calibration, and the execution time per update took 0.013ms on average, with the full series being processed in 12.26 seconds. The code for this data example can be found in Appendix \ref{app:Soundsensing}, but is omitted from the replication materials as the data cannot be shared publicly.


\section{Discussion}\label{sec:discussion}
In this paper we have presented the package \pkg{gridcp}, an online changepoint detection library built on the methodology of \citet{moen_grid}. To close, we discuss three points concerning \pkg{gridcp}'s scope and behavior: a structural assumption that \pkg{gridcp} inherits from the underlying methodology, an open question of calibration over an unbounded horizon, and a design choice in how the detector is reset after an alarm.

First, \pkg{gridcp} requires that the test statistic (the score) can be written in the form of~\eqref{eq:framework-test}. That is, it must be written in terms of running summaries whose per-split score and penalty are computable in constant time. This excludes certain tests, such as those based on order statistics \citep{moen_grid}, which require maintaining and re-sorting the full ordered data sequence when a new observation arrives. We remark that this limitation is inherent to the methodology proposed by \cite{moen_grid}, and not an artifact of the implementation in \pkg{gridcp}. Nonetheless, the admissible class of test statistics is arguably broad: as Section~\ref{sec:builtins} illustrates, it spans CUSUM, likelihood-ratio, regression, nonparametric, and generalized likelihood-ratio tests, covering many of the changepoint scenarios encountered in practice. It is also considerably wider than the model classes supported by the online detectors surveyed in Section~\ref{sec:intro}, many of which are tied to a single parametric problem, such as the Gaussian mean shifts with identity covariance of \code{ocd} \citep{chen_high-dimensional_2022}. Users may, in addition, supply their own tests through the same interface.

Second, the calibration methods in \pkg{gridcp}, and in particular the false alarm probability calibration of Section~\ref{subsec:calib-fa}, are based on null streams of a finite horizon, even though monitoring may continue indefinitely. This reflects a genuine open problem, rather than a shortcoming specific to \pkg{gridcp}, since infinite-horizon calibration appears to have received little attention in the literature. We argue that this is a limited practical concern: real data streams are never truly infinite, and what matters is that the detector can be calibrated to a horizon as long as the intended deployment. Two features of \pkg{gridcp} make this feasible. First, the package is fast enough to calibrate on very long null streams directly. Second, the time-dependent penalty makes the detector progressively more conservative over time, so that the empirical false alarm rate may stay close to the target well beyond the calibration horizon. Calibrating on a comparatively short horizon can therefore suffice for a much longer deployment. This behavior is well supported empirically (see Figure~\ref{fig:fa_rate}) and theoretically \citep{moen_grid} for the CUSUM score. A precise characterization of when, and for which models, a shorter calibration horizon suffices is left to future work.

Third, a design choice that directly affects which changepoints are detected is how the detector is reset after an alarm.  \pkg{gridcp} provides a single reset operation, which re-initializes the detector to its initial state, thus discarding all accumulated summaries and grid points. The implicit default is to apply it whenever an alarm is raised, so that monitoring restarts from scratch at the alarm time.
How best to continue monitoring after an alarm is itself a question that has received limited attention, and our choice is deliberately conservative: by discarding the past entirely, we never risk contaminating the post-reset sample with pre-change data. The price is that the alarm time lags the true change, and consequently, closely spaced changepoints may be skipped, as illustrated by Figure~\ref{fig:well_log}. 
A natural alternative would be to resume from the split point of maximum score, i.e., localize the changepoint. However, estimating the changepoint risks placing it too early, reintroducing the very contamination that the full reset avoids; it also requires backtracking over the previously processed data, which may not be feasible in high-frequency settings. 
We defer a fuller study of reset strategies to future work.

We emphasize that none of these points prevent \pkg{gridcp} from working well in practice: the latter two, in particular, are design choices we have found effective across a range of settings. Overall, \pkg{gridcp} offers a complete pipeline for fast online changepoint detection: a broad library of built-in tests (with the possibility of user-defined ones), along with methods for calibrating the detector to a target false alarm probability or average run length, all backed by established theoretical guarantees. We hope it lowers the barrier to deploying online changepoint detection in practice.





\section*{Acknowledgments}

\begin{leftbar}
We thank Idris Eckley for insightful discussions. This work was supported by the research project SODA (Norwegian Research Council, project number
356322) and the research center Integreat (Norwegian Research Council, project number 332645). Data was provided by Soundsensing. Data collection was partially funded by the project EarOnEdge, which was supported by Norwegian Research Council from 2022-2026.  During the development of the software package, the authors occasionally used Claude Code (Anthropic) to assist with high-level software design, code generation and refactoring. After using this tool, the authors rigorously reviewed, modified, and validated all generated code. The authors take full responsibility for the final implementation.
\end{leftbar}


\bibliography{refs}

\begin{appendix}
\section{Mathematical details of the built-in score models}\label{app:scoremath}
This appendix gives the explicit test statistic computed by each built-in score model of Section~\ref{sec:builtins}, together with the values of $M$ and $\mathrm{df}$ underlying the penalty~\eqref{penaltylm}. Throughout, a split point $b$ divides the observations $Y_1,\dots,Y_t$ into a pre-change segment $Y_{1:(b-1)}$ and a post-change segment $Y_{b:t}$.

\subsection{Further details on the Chi-squared type penalty}
With the exception of \code{NPFOCuS} and \code{RegressionMcScan}, every built-in score model admits a natural chi-squared-type penalty \eqref{penaltylm}, whose parameters $M$ and $\mathrm{df}$ we now specify. Under the null, recall that each such score equals the maximum of $M$ exact or approximate chi-squared variables with $\mathrm{df}$ degrees of freedom, each centered at its null mean. The degrees of freedom $\mathrm{df}$ depend on the score model and aggregation, while $M$ corrects for the coordinate-wise maximum: $M=p$ for a score formed as a coordinate-wise maximum (the \code{"max"} aggregation and the max component of \code{"max-sum"}), and $M=1$ otherwise, including the joint scores. For \code{CUSUM}, for instance, each coordinate-wise squared CUSUM is exactly $\chi^2$ with one degree of freedom before centering, so \code{"max"} gives $M=p$ and $\mathrm{df}=1$, and \code{"sum"} gives $M=1$ and $\mathrm{df}=p$. Both are computed automatically, and the value of $\mathrm{df}$ for each score model follows from the statistic given in its subsection below.

For exactly-chi-squared scores, for example \code{CUSUM} and \code{RegressionWald}, the threshold \newline$\lambda\,\mathrm{pen}_{M,\mathrm{df}}(t)$ controls the false alarm probability over an unbounded horizon whenever the threshold $\lambda$ is chosen sufficiently large. Indeed, for any fixed $t$ and $b$, assume that the score $S_b^{(t)}$ takes the form $S_b^{(t)} = \max_{i=1,\ldots, M}\ (X_i - \mathrm{df})$, where $X_i\overset{\mathrm{i.i.d.}}{\sim} \chi^2(\mathrm{df})$. That is, the score $S_b^{(t)}$ is the maximum of $M$ independent and centered chi-squared random variables with $\mathrm{df}$ degrees of freedom. Then the concentration inequality of \citet[][Lemma 1]{laurentmassart} gives 
$$\PP(X_i - \mathrm{df} \ge 2\sqrt{\mathrm{df}\,x}+2x)
\le e^{-x},$$ for any $x>0$. Fix a target false alarm probability $\delta\in(0,1)$ and take $x=\log\!\bigl(3\delta^{-1}Mt^{2}\log t\bigr)$, so that $Me^{-x}=\delta/(3t^{2}\log t)$. A union bound over the $M$ individual chi-squares then gives $\PP\{S_b^{(t)} \ge 2x+2\sqrt{\mathrm{df}\,x}\} \le \delta/(3t^{2}\log t)$. Since $x \le \{3+\log(3/\delta)\}\log(tM)$ for every $t\ge2$, we have $2x+2\sqrt{\mathrm{df}\,x} \le \lambda\,\mathrm{pen}_{M,\mathrm{df}}(t)$ whenever $\lambda \ge 6+2\log(3/\delta)$, and therefore
\begin{equation}\label{eq:penaltybound}
  \PP_{\infty}\{S_b^{(t)} > \lambda\,\mathrm{pen}_{M,\mathrm{df}}(t)\} \leq \frac{\delta}{3t^2\log t}.
\end{equation} 
This is Assumption 2A in \cite{moen_grid}, and Proposition 4 in the same paper then implies that the stopping rule $\widehat{\tau}$ in \eqref{overalltest} satisfies $\PP_{\infty}(\tauhat <\infty) \leq \delta$. While this theoretical choice of $\lambda$ is generally very conservative, this shows that the penalty function in \eqref{penaltylm} can indeed control the false alarm probability as long as $\lambda$ is chosen sufficiently large. 

\subsection{Gaussian scores}\label{app:gaussmath}
Each Gaussian score is the closed-form likelihood-ratio statistic for a change in the Gaussian model~\eqref{gaussianmodel}, differing in which parameters may change and what is assumed of the nuisance parameters (Table~\ref{tab:gaussian}).

\paragraph[CUSUM]{The \code{CUSUM} score model.}
The \code{CUSUM} score model tests for a change in the mean $\mu$ with the covariance known and equal to the identity. The per-coordinate score is the squared CUSUM $\bigl(C_b^{(t)}(j)\bigr)^2$, where $C_b^{(t)}(j)$ is coordinate $j$ of the CUSUM vector~\eqref{CUSUMex}; it is exactly $\chi^2(1)$ under the null, and the $p$ coordinates are combined depending on the choice of aggregation, explained in Section~\ref{sec:builtins}. For example, \code{aggregation = "max-sum"} returns
\begin{equation}\label{eq:cusum-maxsum}
  S_b^{(t)} = \begin{pmatrix}
    \max_{1\le j\le p}\bigl(C_b^{(t)}(j)\bigr)^2 - 1 \\[2pt]
    \sum_{j=1}^{p}\bigl(C_b^{(t)}(j)\bigr)^2 - p
  \end{pmatrix},
\end{equation}
that is, the max and the sum across all coordinates, centered by subtracting their respective $\chi^2$ null means $1$ and $p$. A split point with an empty segment returns $0$.

\paragraph[GaussianMean]{The \code{GaussianMean} score model.}
The \code{GaussianMean} score model tests for a change in the mean $\mu$ with the covariance unknown but constant across the change. The covariance is estimated in one of two ways, depending on the assumed form of the covariance matrix (diagonal or full), selected by the input parameter \code{cov\_estimate}. With \code{cov\_estimate = "diagonal"} (the default), the covariance is taken to be diagonal and the score is computed per coordinate. For each coordinate $j$, the score is the profile log-likelihood ratio for a change in mean
\citep[Section~2.1]{chen2012parametric},
\begin{equation}\label{eq:gaussmean-diag}
  S_b^{(t)}(j) = t\bigl(\log\widehat\sigma^2_{1:t}(j)
  - \log\widehat\sigma^2_{1:(b-1),\,b:t}(j)\bigr) - 1.
\end{equation}
Here,
\begin{align}\label{eq:block-moments}
  \bar Y_{a:b}(j) &= (b-a+1)^{-1}\sum_{i=a}^{b} Y_i(j), &
  \widehat\sigma^2_{a:b}(j) &= (b-a+1)^{-1}\sum_{i=a}^{b}\bigl(Y_i(j)-\bar Y_{a:b}(j)\bigr)^2,
\end{align}
are the block sample mean and variance for the $j$-th coordinate of block $Y_{a:b}$, and
\begin{equation}\label{eq:pooled-variance}
  \widehat\sigma^2_{1:(b-1),\,b:t}(j) = \frac1t\Bigl[
    \sum_{i=1}^{b-1}\bigl(Y_i(j)-\bar Y_{1:(b-1)}(j)\bigr)^2
    + \sum_{i=b}^{t}\bigl(Y_i(j)-\bar Y_{b:t}(j)\bigr)^2\Bigr]
\end{equation}
is the pooled sample variance across the pre- and post-change segments. The un-centered statistic is asymptotically $\chi^2(1)$ \citep{wilks1938large}, and the per-coordinate scores are aggregated based on the choice of \code{aggregation}, as explained in Section~\ref{sec:builtins}.

With \code{cov\_estimate = "full"}, the covariance is estimated jointly across coordinates and the score is a single joint statistic, that is, the (centered) log-likelihood ratio for a change in mean with a pooled sample covariance \citep[Section~3.1]{chen2012parametric},
\begin{equation}\label{eq:gaussmean-full}
  S_b^{(t)} = t\bigl(\log\det\widehat\Sigma_{1:t}
  - \log\det\widehat\Sigma_{1:(b-1),\,b:t}\bigr) - p,
\end{equation}
which is asymptotically $\chi^2(p)$ un-centered, where $\widehat\Sigma_{a:b}$ is the block sample covariance and $\widehat\Sigma_{1:(b-1),\,b:t}$ is pooled across the two segments as in~\eqref{eq:pooled-variance}. The diagonal score requires at least three observations in each segment; the full score requires $t \ge 2p+2$ and returns $0$ otherwise.

\paragraph[GaussianVariance]{The \code{GaussianVariance} score model.}
The \code{GaussianVariance} score model tests for a change in the variance with the mean known to be zero and the covariance diagonal, so the score is computed per coordinate. For each coordinate $j$, the score is the log-likelihood ratio for a change in variance \citep[Section~2.2]{chen2012parametric},
\begin{equation}\label{eq:gaussvar}
  S_b^{(t)}(j) = t\log\widehat\sigma^2_{1:t}(j)
  - (b-1)\log\widehat\sigma^2_{1:(b-1)}(j)
  - (t-b+1)\log\widehat\sigma^2_{b:t}(j) - 1.
\end{equation}
Since the mean is assumed to be zero, the variance is estimated without mean-centering,
\begin{equation}\label{eq:gaussvar-moment}
  \widehat\sigma^2_{a:b}(j) = (b-a+1)^{-1}\sum_{i=a}^{b} Y_i(j)^2,
\end{equation}
and the (un-centered) statistic is asymptotically $\chi^2(1)$. A split point with an empty segment returns $0$. The per-coordinate scores are aggregated based on the choice of \code{aggregation}, as explained in Section~\ref{sec:builtins}. A nonzero mean should be subtracted before the observations are passed to this score.

\paragraph[GaussianMeanOrVariance]{The \code{GaussianMeanOrVariance} score model.}
The \code{GaussianMeanOrVariance} score model tests for a change in the mean and/or variance, both unknown and the covariance diagonal, and the score is computed per coordinate. For each coordinate $j$, the score is
\citep[Section~2.3]{chen2012parametric}
\begin{equation}\label{eq:meanorvar}
  S_b^{(t)}(j) = t\log\widehat\sigma^2_{1:t}(j)
  - (b-1)\log\widehat\sigma^2_{1:(b-1)}(j)
  - (t-b+1)\log\widehat\sigma^2_{b:t}(j) - 2,
\end{equation}
where $\widehat\sigma^2_{a:b}(j)$ is the mean-centered sample variance~\eqref{eq:block-moments}. It is asymptotically $\chi^2(2)$ (un-centered), with one degree of freedom for the mean and one for the variance. The per-coordinate scores are aggregated based on the choice of \code{aggregation}, as explained in Section~\ref{sec:builtins}. The score requires at least three observations in each segment, returning 0 otherwise.

\paragraph[GaussianMeanOrCovariance]{The \code{GaussianMeanOrCovariance} score model.}
The \code{GaussianMeanOrCovariance} score model tests for a change in the mean and/or covariance, both unknown and the covariance full, and the score is a single joint statistic. Using the sample covariance in place of the per-coordinate variance \citep[Section~3.3]{chen2012parametric}, the score is
\begin{equation}\label{eq:meanorcov}
  S_b^{(t)} = t\log\det\widehat\Sigma_{1:t}
  - (b-1)\log\det\widehat\Sigma_{1:(b-1)}
  - (t-b+1)\log\det\widehat\Sigma_{b:t} - \mathrm{df},
\end{equation}
asymptotically $\chi^2(\mathrm{df})$ (un-centered), where $\widehat\Sigma_{a:b}$ is the block sample covariance and $\mathrm{df} = p + p(p+1)/2$ is the number of free mean and covariance parameters. The score returns $0$ unless each segment contains more than $2p$ observations, and also when a segment covariance or the pooled covariance is singular.

\subsection{Regression scores}\label{app:regmath}

Both score models concern the linear model in~\eqref{regressionmodel}, 
with $X_i \in \RR^q$ and a change from $\beta_1$ to $\beta_2 \neq \beta_1$ at the
changepoint; they differ in their assumptions on the covariates and the noise.
\code{RegressionWald} treats the covariates as fixed and the noise as Gaussian,
$\epsilon_i \sim \N(0,1)$. Writing $M_k=\sum X_iX_i^\top$ and $S_k=\sum Y_iX_i$ for the
Gram matrix and cross-product sum over segment $k$, and
$\widehat\beta_k = M_k^{-1}S_k$ for the least-squares estimate, the score is the
two-sample Wald contrast
\begin{equation}\label{eq:regwald}
  S_b^{(t)} = \bigl(\widehat\beta_1-\widehat\beta_2\bigr)^\top
  \bigl(M_1^{-1}+M_2^{-1}\bigr)^{-1}
  \bigl(\widehat\beta_1-\widehat\beta_2\bigr) - q ,
\end{equation}
which is exactly $\chi^2(q)$ under the null (un-centered); it returns
$0$ unless each segment has at least $q$ observations, so that $M_1$ and $M_2$ are
of full rank. \code{RegressionMcScan} instead treats the covariates as random, with
$\EE[X_i]=0$ and $\EE[X_iX_i^\top]=\Sigma$, and for $b\in\{2,\dots,t\}$ uses the
``McScan'' statistic of \cite{cho2025detection},
\begin{equation}\label{eq:mcscan}
  S_b^{(t)} = \sqrt{\frac{(b-1)(t-b+1)}{t}}\,
  \max_{1\le j\le q}\ \left|\frac1{b-1}\sum_{i=1}^{b-1}Y_iX_i(j)
  - \frac1{t-b+1}\sum_{i=b}^{t}Y_iX_i(j)\right|,
\end{equation}
with penalty $\mathrm{pen}(t)=\sqrt{\log(qt)}$, motivated by Theorem~2 of
\citet{cho2025detection} under their Assumptions~1 and~2(i).

\subsection[ExponentialFamilyGLR]{The \code{ExponentialFamilyGLR} score model}\label{app:glrmath}

The \code{ExponentialFamilyGLR} score model tests for a change in the natural parameter $\theta$ of the exponential family~\eqref{eq:expfam}. The score is the centered generalized log-likelihood ratio for such a change \citep[see, e.g.,][]{lai2010sequential},
\begin{equation}\label{eq:glr}
  S_b^{(t)} = 2\bigl(\ell(\widehat\theta_{1:(b-1)}) + \ell(\widehat\theta_{b:t})
  - \ell(\widehat\theta_{1:t})\bigr) - v,
\end{equation}
where $\ell(\widehat\theta_{a:c}) = \widehat\theta_{a:c}^\top\sum_{i=a}^{c}h(Y_i)
- (c-a+1)A(\widehat\theta_{a:c})$ is the maximized log-likelihood over the segment
and $\widehat\theta_{a:c}$ is the maximum likelihood estimate of the natural
parameter, obtained by solving the score equation $\nabla A(\widehat\theta_{a:c})
= (c-a+1)^{-1}\sum_{i=a}^{c}h(Y_i)$ with Newton's method. The un-centered
statistic is approximately $\chi^2(v)$ under the null. When there are fewer than
\code{min\_seg} observations in either the pre- or post-change segment, the score model returns $0$, where \code{min\_seg}
defaults to $v+1$.

\paragraph{Specifying the family.}
\code{ExponentialFamilyGLR} requires the user to specify the exponential family, either through a built-in family via the \code{from\_family} class method (Table~\ref{tab:expfam-families}),
\begin{minted}[]{python}
score = ExponentialFamilyGLR.from_family("poisson")
\end{minted}
or by supplying $h$, $A$, and the derivatives of $A$ directly: \code{A\_prime} and \code{A\_dprime} for one-parameter families ($v=1$), or \code{A\_grad} and \code{A\_hess} for multiparameter families ($v>1$). For example, the Poisson distribution with rate $\mu>0$ has natural parameter $\theta=\log\mu$, sufficient statistic $h(x)=x$, and log-partition function $A(\theta)=e^\theta$, and the corresponding score is defined as follows:
\begin{minted}[linenos, frame=lines]{python}
@nb.njit
def h(x):
    return x[0]
@nb.njit
def A(theta):
    return np.exp(theta)
@nb.njit
def A_prime(theta):
    return np.exp(theta)
@nb.njit
def A_dprime(theta):
    return np.exp(theta)
score = ExponentialFamilyGLR(
    v = 1, n_features = 1,
    h = h, A = A, A_prime = A_prime, A_dprime = A_dprime,
)
\end{minted}

At construction time the score builds MLE solvers and GLR kernels tailored to the family. When all user-supplied callables are \pkg{Numba}-compiled these are JIT-compiled and run with no \proglang{Python} overhead, and otherwise fall back to plain \pkg{NumPy}; numerical stability is ensured through regularized Hessians and a backtracking line search, and parameter constraints may be imposed via \code{theta\_min} and \code{theta\_max}.

\subsection[NPFOCuS]{The \code{NPFOCuS} score model}\label{app:npfocusmath}

For a fixed value grid $\mathcal U=\{u_1,\dots,u_m\}$, the sufficient statistic at coordinate $j$ is
$h(Y_i(j))=(\ind\{Y_i(j)\le u_1\},\dots,\ind\{Y_i(j)\le u_m\})$. With the maximized
Bernoulli log-likelihood
\begin{equation}
  \ell(k;n) = \begin{cases}
    k\log(k/n) + (n-k)\log(1-k/n), & 0<k<n,\\
    0, & k\in\{0,n\},
  \end{cases}
\end{equation}
the likelihood ratio statistic for a change at $b$ in the threshold-$u_k$
indicator is
\begin{equation}
  \begin{aligned}
    \Lambda_{b,k}(j) = 2\Bigl[
    &\ \ell\bigl(\textstyle\sum_{i=1}^{b-1}\ind\{Y_i(j)\le u_k\};\,b-1\bigr)
    + \ell\bigl(\textstyle\sum_{i=b}^{t}\ind\{Y_i(j)\le u_k\};\,t-b+1\bigr)\\
    &\ - \ell\bigl(\textstyle\sum_{i=1}^{t}\ind\{Y_i(j)\le u_k\};\,t\bigr)\Bigr].
  \end{aligned}
\end{equation}
For each coordinate $j$, the $\Lambda_{b,k}(j)$ are aggregated over the value grid both by
summing and by maximizing, giving the score
\begin{equation}
  S^{(t)}_b(j) = \begin{pmatrix}
    \sum_{k=1}^{m}\Lambda_{b,k}(j)\\[4pt]
    \max_{1\le k\le m}\Lambda_{b,k}(j)
  \end{pmatrix}.
\end{equation}
For multivariate data the coordinate-wise scores are combined in accordance with the \code{aggregation} argument. Unlike the chi-squared scores, \code{NPFOCuS} carries no time-dependent penalty. A split point with an empty segment returns $0$.

    \section{Building a custom score model from scratch}
This appendix explains how to implement a custom score, i.e. a custom changepoint test statistic. To be compatible with \pkg{gridcp}, the score must be implemented as a \emph{score model}: a class that contains methods for computing running summary statistics and split-wise scores used by \code{GridDetector}; specifically, the class and its methods must be defined in accordance with the \code{ScoreModel} protocol.
Once a custom score model is compatible with this protocol, it can be plugged into \code{GridDetector} without modification. Section~\ref{subsec:scoremodel} summarizes the \code{ScoreModel} contract, and Section~\ref{customscoresubsec} gives a minimal and univariate implementation of the built-in score model \code{CUSUM} described in Section \ref{sec:builtins}.

\subsection[The ScoreModel protocol]{The ScoreModel protocol}\label{subsec:scoremodel}
The \code{ScoreModel} protocol functions as a contract between the user and the \code{GridDetector}: a class is compatible with the \code{ScoreModel} protocol whenever it provides the required methods and attributes. In particular, a compatible class is not required to inherit from any particular base class. Once a custom score model is compatible with the \code{ScoreModel} protocol, the \code{GridDetector} will accept it without explicit inheritance from \code{ScoreModel}, and so will the utility functions in Section~\ref{sec:calibration}. 

A simplified version of the \code{ScoreModel} protocol (omitting the \code{@runtime\_checkable} decorator, which enables \code{isinstance} checks at runtime) is given below:
\begin{minted}[linenos, frame=lines]{python}
class ScoreModel(Protocol[TScoreState]):
    @property
    def n_features(self) -> int: ...
    @property
    def n_scores(self) -> int: ...
    def init_state(self) -> TScoreState: ...
    def update(self, state: TScoreState, x: ArrayLike) -> TScoreState: ...
    def compute_penalized_scores(
        self,
        state: TScoreState,
        grid_states: list[TScoreState],
    ) -> np.ndarray: ...
\end{minted}
Here, \code{TScoreState} is any user-specified container for a running summary $H_j$ from \eqref{summary}. The protocol does not require any particular storage format, and is intentionally flexible: \code{TScoreState} may be any immutable snapshot that carries the information needed to compute a score, such as a frozen dataclass, a named tuple, or a plain tuple. 

The role of each protocol member is as follows.
\begin{itemize}
  \item \code{n\_features} and \code{n\_scores} are read as plain attributes, written \code{score.n\_features} and \code{score.n\_scores}. They may be implemented either as class-level attributes (as in Section~\ref{customscoresubsec}) or as \code{@property} methods; both satisfy the protocol. \code{n\_features} gives the observation dimension $p$; \code{n\_scores} gives the number of scores produced at a given split point, allowing a single implementation to output multiple scores.
  \item \code{init\_state()} returns an empty \code{TScoreState} object.
 \item The method \code{update(state, x)} advances the summary from $H_{j-1}$ (\code{state}) to $H_j$ after observing $Y_j$ (\code{x}), and returns a new \code{TScoreState} object. The input \code{state} must not be mutated in place, because \code{GridDetector} stores it as an immutable snapshot at the corresponding split point.
    \item The method \code{compute\_penalized\_scores(state, grid\_states)} computes and returns penalized scores $\{S_b^{(t)}/\mathrm{pen}(t)\}_{b\in B^{(t)}}$ from the running summary $H_t$ (\code{state}) and a list of summaries $\{H_{b-1}\}_{b \in B^{(t)}}$ (\code{grid\_states}). The output is a matrix of dimension $G \,\times$ \code{n\_scores}, where $G = |B^{(t)}|$. Row $g$ corresponds to one specific split point $b$, and column $k$ to the $k$-th score. Even a single-score model must return a two-dimensional array of shape $(G, 1)$, not a one-dimensional array.
\end{itemize}

We remark that the value of \code{n\_scores} must remain constant across calls, because it fixes both the detector threshold width in the \code{GridDetector} and the number of score columns expected at each time step. Moreover, \code{compute\_penalized\_scores} must always return a two-dimensional array whose second dimension matches that declared \code{n\_scores} value.

\subsection{Implementing a custom score model}\label{customscoresubsec}
The code below implements the \code{ScoreModel} protocol for the univariate CUSUM score based on the squared CUSUM of~\eqref{CUSUMex}. The score model (\code{CusumScore}) is stateless; all running statistics are held in the companion state class (\code{CusumState}). Keeping the score model stateless and the state container immutable is recommended practice for robust implementations, but the protocol requires neither a particular score class structure nor a particular state type.

\begin{minted}[linenos, frame=lines]{python}
from dataclasses import dataclass
import numpy as np
from gridcp import GridDetector
@dataclass(frozen = True, slots = True)
class CusumState:
    n: int = 0
    s: float = 0.0
class CusumScore:
    n_features = 1
    n_scores = 1
    def init_state(self):
        return CusumState()
    def update(self, state, x):
        v = float(np.asarray(x).ravel()[0])
        return CusumState(n = state.n + 1, s = state.s + v)
    def compute_penalized_scores(self, state, grid_states):
        t = state.n
        penalty = np.log(t) + np.sqrt(np.log(t))
        scores = np.empty((len(grid_states), 1))
        for i, gs in enumerate(grid_states):
            n_pre = gs.n
            s_pre = gs.s
            n_post = t - n_pre
            s_post = state.s - s_pre
            if n_pre > 0 and n_post > 0:
                cusum = (np.sqrt(n_post / (t * n_pre)) * s_pre
                         - np.sqrt(n_pre / (t * n_post)) * s_post)
                scores[i, 0] = (cusum**2 - 1.0) / penalty
            else:
                scores[i, 0] = 0.0
        return scores
detector = GridDetector(score = CusumScore(), threshold = 1.0)
\end{minted}

Above, the class \code{CusumState} is the choice of \code{TScoreState} for the custom CUSUM score model. The \code{CusumState} is a container storing two entries: the time-stamp \code{n}, giving the number of samples, and \code{s}, the cumulative sum $\sum_{i=1}^n Y_i$ at this sample size. Decorating it with \code{@dataclass(frozen=True, slots=True)} serves two purposes: \code{frozen=True} makes instances immutable, preventing accidental in-place mutation (which would silently corrupt the stored grid snapshots), and \code{slots=True} reduces per-instance memory overhead. Inside \code{compute\_penalized\_scores}, the input argument \code{state} plays the role of $H_t$ and each element of \code{grid\_states} plays the role of one pre-change summary $H_{b-1}$. Subtracting a stored snapshot from the current totals yields the post-change count $n_2 = t - n_1$ and cumulative sum, from which the squared CUSUM statistic follows directly; the centering term $-1$ subtracts the $\chi^2(1)$ null mean of the squared CUSUM~\eqref{CUSUMex}. Although only one (penalized) score is produced, the result is returned as a $(G, 1)$ matrix, where $G$ is the length of \code{grid\_states}; moreover, the time-dependent penalty $\log t + \sqrt{\log t}$ is derived from \code{state.n}, which holds the overall sample size $t$. The built-in \code{CUSUM} score model class implements the same score with a \pkg{Numba}-compiled inner loop for higher throughput, and moreover accepts multivariate inputs.

For further details and method signatures, we refer to the documentation of \pkg{gridcp}. 

    \section{Additional code examples}
\subsection{Code for analyzing soundlevel data}\label{app:Soundsensing}
The below code demonstrates how to apply a Gaussian change-in-mean detector (variance unknown) to the data provided by Soundsensing AS. 
\begin{minted}[linenos, frame=lines]{python}
import pandas as pd
from pathlib import Path
import pyarrow.parquet as pq
from gridcp.scores import GaussianMean
from gridcp.calibration import calibrate_threshold_false_alarm_from_data
import time
from gridcp.detector import GridDetector

single_path = Path('data.parquet')
data = pq.read_table(single_path, read_dictionary=[]).to_pandas()

device_id = data.index.get_level_values('device_id').unique()[3]
device_data = data.xs(device_id, level='device_id')['leq'].dropna()

start1 = pd.Timestamp('2024-05-5', tz=device_data.index.tz)
end1 = pd.Timestamp('2024-05-6', tz=device_data.index.tz)

training_series = device_data[
    ((device_data.index >= start1) & (device_data.index < end1))
]
full_series = device_data

score = GaussianMean(n_features=1, enable_penalty=True)
threshold = calibrate_threshold_false_alarm_from_data(
    score,
    training_series,
    stream_len = 200, 
    false_alarm_probability=0.001,
    n_paths=20000,
    rng=42,
)

alarm_times = []
t_start = time.perf_counter()
for t, x in enumerate(device_data.values):
    state, output = detector.update(state, x)
    if output["alarm"]:
        alarm_times.append(t)
        state = detector.init_state() 
t_end = time.perf_counter()

print(f"Alarms at indices: {alarm_times}")
alarm_timestamps = device_data.index[alarm_times]
print(f"Alarm timestamps: {list(alarm_timestamps)}")
print(f"Runtime for full time series ({len(device_data)} samples): 
t_end - t_start:.3f} s")

\end{minted}

\end{appendix}

\newpage

\end{document}